\documentclass{sig-alternate}  
\usepackage{graphicx}
\usepackage{amssymb}
\usepackage{amsmath}
\usepackage{times}
\usepackage{multicol,multirow}
\usepackage{epsf,psfrag,pstricks}
\usepackage{mdwlist}
\usepackage{rotating}
\usepackage{xspace}
\usepackage{url}
\usepackage{capt-of}
\usepackage[ruled,vlined]{algorithm2e}
\usepackage[noend]{algpseudocode}

\newcommand{\cate}{\textsf{CaTE}\xspace}

\newcommand{\cref}[1]{Chapter~\ref{#1}}

\newcommand{\ie}{i.\,e., \@}
\newcommand{\eg}{e.\,g., \@}

\newcommand{\etal}{et~al.\xspace}

\newcounter{fn1}
\newcounter{fn2}
\newcounter{fn3}
\newcounter{fn4}
\newcounter{fn5}
\begin{document}

\title{Content-aware Traffic Engineering}

\numberofauthors{5}
\author{
\alignauthor Benjamin Frank\\
        \affaddr{T-Labs/TU Berlin} \\
        \affaddr{\small{bfrank@net.t-labs.tu-berlin.de}}
\alignauthor Ingmar Poese\\
        \affaddr{T-Labs/TU Berlin} \\
        \affaddr{\small{ingmar@net.t-labs.tu-berlin.de}}
\and
\alignauthor Georgios Smaragdakis\\
        \affaddr{T-Labs/TU Berlin} \\
        \affaddr{\small{georgios@net.t-labs.tu-berlin.de}}
\alignauthor Steve Uhlig\\
        \affaddr{\mbox{Queen Mary University of London}} \\
        \affaddr{\small{steve@eecs.qmul.ac.uk}}
\alignauthor Anja Feldmann\\
       \affaddr{T-Labs/TU Berlin} \\
       \affaddr{\small{anja@net.t-labs.tu-berlin.de}}
}

\maketitle

\begin{abstract}\label{abstract}

Today, a large fraction of Internet traffic is originated by Content Providers
(CPs) such as content distribution networks and hyper-giants.
To cope with the increasing demand for content, CPs deploy massively distributed infrastructures.
This poses new challenges for CPs as they have to dynamically map end-users to appropriate
servers, without being fully aware of network conditions within an ISP as well as the end-users 
network locations. Furthermore, ISPs struggle to cope with rapid traffic shifts caused by the 
dynamic server selection process of CPs.

In this paper, we argue that the challenges that CPs and ISPs face separately today can be turned 
into an opportunity. We show how they can jointly take advantage of the deployed distributed 
infrastructures to improve their operation and end-user performance. We propose 
\emph{Content-aware Traffic Engineering} (\cate), which dynamically adapts the traffic 
demand for content hosted on CPs by utilizing ISP network information and end-user location 
during the server selection process. As a result, CPs enhance their end-user to server mapping 
and improve end-user experience, thanks to the ability of network-informed server selection to 
circumvent network bottlenecks. In addition, ISPs gain the ability to partially influence the traffic 
demands in their networks. Our results with operational data show improvements in path length 
and delay between end-user and the assigned CP server, network wide traffic reduction of up to 
$15\%$, and a decrease in ISP link utilization of up to $40\%$ when applying \cate to traffic 
delivered by a small number of major CPs.

\end{abstract}

\section{Introduction}\label{sec:Introduction}

People value the Internet for the content it makes available~\cite{CCN}. For example, the 
demand for online entertainment and web browsing has exceeded 70\% of the peak downstream 
traffic in the United States~\cite{sandvine11}. 
Recent traffic studies~\cite{TrafficTypesGrowth:2011,arbor,PADIS2010} show that a large fraction 
of Internet traffic is originated by a small number of Content Providers (CPs). Major CPs are highly 
popular rich media sites like YouTube and Netflix, One-Click Hosters (OCHs), \eg RapidShare or 
MegaUpload, as well as Content Delivery Networks (CDN) such as Akamai or Limelight and
hyper-giants, \eg Google, Yahoo! or Microsoft. Gerber and Doverspike~\cite{TrafficTypesGrowth:2011} 
report that a few CPs account for more than half of the traffic of a US-based Tier-1 carrier. 
Poese \etal~\cite{PADIS2010} report a similar observation from the traffic of a European 
Tier-1 carrier. Labovitz \etal~\cite{arbor} infer that more than 10\% of the total Internet 
inter-domain traffic originates from Google, and Akamai claims to deliver more than 20\% 
of the total Web traffic in the Internet~\cite{Akamai-Network}. In North America, Netflix 
is responsible for around 30\% of the traffic during peak hours~\cite{sandvine11} by offering 
a high definition video streaming service hosted on CDN infrastructures such as Limelight and 
the CDN operated by Level3.

To cope with the increasing demand for content, CPs deploy massively distributed server 
infrastructures~\cite{ImprovingPerformanceInternet2009} to replicate content and make it 
accessible from different locations in the Internet~\cite{CDNsec2009,Cartography}. For example, 
Akamai operates more than $60,000$ servers in more than $5,000$ locations across nearly $1,000$
networks~\cite{ImprovingPerformanceInternet2009,Akamai-Network}. Google is reported to operate 
tens of data-centers and front-end server clusters worldwide~\cite{MovingBeyondE2E2009,Tariq:What-if}.
Microsoft has deployed its CDN infrastructure in 24 locations around the world~\cite{WindowsAzure}. Amazon maintains
at least 5 large data-centers and caches in at least 21 locations around the world~\cite{amazon}. Limelight operates
thousands of servers in more than 22 delivery centers and connects directly to more than 900 networks worldwide~\cite{LLNetworks}.

The growth of demand for content and the resulting deployment of content delivery infrastructures pose 
new challenges to CPs and to ISPs. For CPs, the cost of deploying and maintaining such a massive 
infrastructure has significantly increased during the last years~\cite{AkamaiCutting:2009} and the revenue 
from delivering traffic to end-users has decreased due to the intense competition. Furthermore, CPs struggle 
to engineer and manage their infrastructures, replicate content based on end-user demand, and assign users
to appropriate servers.

The latter is challenging as end-user to server assignment is based on inaccurate end-user location
information~\cite{Precise:Mao2002,DNS-extension-IP-client}, and inferring the network conditions 
within an ISP without direct information from the network is difficult. Moreover, due to highly distributed 
server deployment and adaptive server assignment, the traffic injected by CPs is volatile. For example, if 
one of its locations is overloaded, a CP will re-assign end-users to other locations, resulting in large traffic shifts 
in the ISP network within minutes. Current traffic engineering by ISP networks adapts the routing and 
operates on time scales of several hours, and is therefore too slow to react to rapid traffic changes caused 
by CPs.

\begin{figure}[t]
  \center\includegraphics[width=0.8\linewidth]{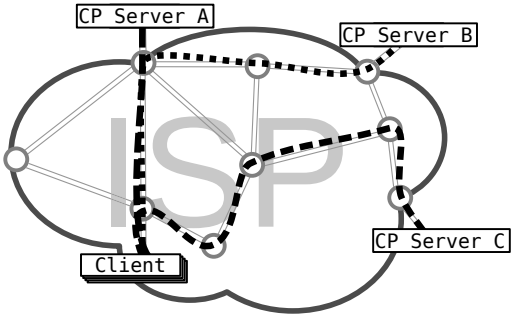}
  \caption{By choosing a CP server for a client with the help of \cate, traffic
  engineering goals and accurate end-user server assignment become possible.}
  \label{fig:flowSelection}
  \vspace{-1.5em}
\end{figure}

The pressure for cost reduction and customer satisfaction that both CPs and ISPs are confronted 
with, coupled with the opportunity that distributed server infrastructures offer, motivate us to 
propose a new tool in the traffic engineering landscape. We introduce \emph{Content-aware 
Traffic Engineering} (\cate). \cate leverages the location diversity offered by CPs and, through 
this, it allows to adapt to traffic demand shifts. In fact, \cate relies on the observation that by 
selecting an appropriate server among those available to deliver the content, the path of the 
traffic in the network can be influenced in a desired way. Figure~\ref{fig:flowSelection} illustrates 
the basic concept of \cate. The content requested by the client is in principle available 
from three servers (A, B, and C) in the network. However, the client only connects to one of the 
network locations. Today, the decision of where the client will connect to is solely done by the 
CP and is partially based on measurements and/or inference of network information and end-user 
location. With \cate the decision on end-user to server assignment can be done jointly between the 
CP and ISP.

\cate complements the existing traffic engineering ecosystem by focusing on traffic demands
rather than routing, by combining (i) the knowledge of CPs about their location diversity and 
server load, with (ii) the ISPs detailed knowledge of the network conditions and end-user location.
\cate offers additional traffic engineering capabilities to ISPs to better manage the volatility 
of CP traffic. Also, thanks to the information about ISP networks, CPs gain the ability to better
assign end-users to their servers and better amortize the cost of deploying and maintaining their 
infrastructure. Furthermore, the burden of measuring and inferring network topology and state 
is removed from the CPs. In short, all involved parties, including the end-users, benefit from 
\cate, creating a win-win situation for everyone. Our contributions are as follows:

\vspace{-0.05in}
\begin{itemize*}
\item We introduce the concept of \cate.
\item We present the design, incentives, and possible deployment schemes of systems to realize \cate.
\item We propose an online algorithm to map end-user requests to servers for \cate and discuss its
properties.
\item We evaluate the performance of \cate using real data from a European Tier-1 ISP. We show that \cate
can improve the assignment of end-users to servers for a number of metrics, namely, link utilization, path length and
path delay. Our results show that the maximum link utilization can be reduced by half, especially during the peak hour,
that the total traffic that flows in the network can be reduced by up to 15\%, and the delay by 20\% respectively
when applying \cate to a small number of major CPs. Similar results are obtained when evaluating \cate on two other
operational networks.
\end{itemize*}
\vspace{-0.05in}

\noindent The remainder of this paper is structured as follows.
In Section~\ref{sec:Background} we present the observations that motivate our work.
In Section~\ref{sec:Approach} we introduce our concept of \cate and present the general architecture as well as
possible deployment sche\-mes. 
We formally define and model \cate in Section~\ref{sec:CaTE}.
We propose algorithms to enable \cate in Section~\ref{sec:Algorithm}.
We evaluate the benefits of \cate in Section~\ref{sec:Evaluation} using data 
from operational networks with different metrics, including link utilization, path delay and length.
We present related work in Section~\ref{sec:Related-Work} and summarize in Section~\ref{sec:Conclusion}.

\section{Challenges and Opportunities in Content Distribution}\label{sec:Background}

With the emergence of ``hyper-giants'' and other popular CPs, the traffic
of the Internet has undergone drastic changes \cite{arbor}.
These changes stem from trends in business and organizational
integration and consolidation. As a consequence, a small number of CPs are responsible for
a large fraction of traffic~\cite{TrafficTypesGrowth:2011,PADIS2010}.
Content delivered by CPs, including highly popular rich media sites like Facebook and high definition video 
streaming such as Netflix or YouTube, is mostly carried over HTTP.
Recent studies unveil that HTTP contributes more than 60\% of
Internet traffic~\cite{UGCcacheability,network-caching,TrafficTypesGrowth:2011,sandvine11,arbor,OnDominantCharacteristics2009}.

Moreover, CPs peer directly with a large number of ISPs and in many locations.
For scalability reasons, most CPs make the content available from all their infrastructure locations~\cite{CDNsec2009}.
The globally deployed infrastructures allow CPs to rapidly shift large amounts of
traffic from one peering point to another. While 
the diverse footprint of CPs and the ability to shift
traffic in short timescales poses new challenges to both CPs and ISPs, 
it also offers new opportunities for joint optimization of content delivery.

\subsection{Challenges in Content Delivery}\label{sec:Challenges}

The scale and complexity of content delivery, especially from distributed infrastructures,
brings multiple challenges to CPs. These challenges have a major impact on both the
end-user performance and ISP operation.

\noindent {\bf Content Delivery Cost.~} CPs strive to minimize the overall cost of delivering huge
amounts of content to end-users. To that end, their assignment strategy is mainly driven by
economic aspects such as bandwidth or energy cost~\cite{AkamaiCutting:2009,Optimizing:Goldenberg2004}.
While a CP will try to assign end-users in such a way that the server can deliver reasonable performance,
this does not always result in end-users being assigned to the server able to deliver the best performance.
Moreover, the intense competition in the content delivery market has led to diminishing returns of
delivering traffic to end-users.

\noindent {\bf End-user Mis-location.~} End-user mapping requests received by the CP DNS servers originate
from the DNS resolver of the end-user, not from the end-user itself. The assignment is therefore based
on the assumption that end-users are close to their DNS resolvers. Recent studies have shown that in
many cases this assumption does not hold~\cite{DNS-IMC-2010,Precise:Mao2002}. As a result, the
end-user is mis-located and the server assignment is not optimal. As a response, DNS extensions have
been proposed to include the end-user IP information~\cite{DNS-extension-IP-client}.

\noindent {\bf Network Bottlenecks.~} Despite their efforts to discover the paths between the end-users
and their servers to predict performances~\cite{sureroute}, CPs have limited information about 
the actual network conditions. Tracking the ever changing network conditions, \ie through
active measurements and end-user reports, incurs an extensive overhead for the CP without a guarantee of performance
improvements for the end-user. Without sufficient information about the network paths between the CP servers
and the end-user, an assignment performed by the CP can lead to additional load on existing network
bottlenecks, or create new ones.

\noindent {\bf End-user Performance.~} Applications delivered by CPs often have requirements in terms of
end-to-end delay~\cite{MovingBeyondE2E2009}. Moreover, faster and more reliable content delivery results 
in higher revenues for e-commerce applications~\cite{Akamai-Network} as well as
user engagement~\cite{Conviva2011}. Despite the significant efforts of CPs, end-user mis-location and
the limited view of network bottlenecks are major obstacles to improve end-user performance.

\subsection{Opportunities for \textbf{\cate}}\label{sec:Opportunities}

The idea behind \cate is to provide solutions for the new challenges in content delivery. 
Indeed, ISPs are in a unique position, both in terms of knowledge as well as incentives, 
to improve content delivery. ISPs have the knowledge about the state of the underlying 
network topology and the status of individual links. This information not only helps CPs 
in their user-to-server mapping, but also reduces the need for CPs to perform large-scale 
active measurements and topology discovery \cite{sureroute}. It also enables CPs 
to better amortize their existing infrastructure, offer better quality of experience to their 
users, and postpone their infrastructure expansion.

The opportunity for ISPs to coordinate with CPs in their server selection is technically possible 
thanks to the decoupling of the server selection from the content delivery. In general, 
any end-user requesting content from a CP first does a mapping request, usually through the 
Domain Name System (DNS). During this request the CP needs to locate the network position 
of the end-user and assign a server capable of delivering the content, preferably close to
the end-user. However, locating the user in a network and inferring the conditions of the path between 
the end-user and eligible CP servers is hard as the CP is missing network information. In contrast, 
ISPs have this information ready at their fingertips, but are currently missing a communication
channel to inform the CPs. Furthermore, ISPs face the challenge of predicting the CP traffic, which
is very difficult due to the lack of information on the mapping of end-users to server decided by CPs.

We propose to use \cate during the server selection process of CPs. In today's 
CP deployment, the server selection is done directly between the end-user and the CP 
without the involvement of the ISP (see arrow A in Figure~\ref{fig:system-overview}). 
Through \cate, CPs are offered the 
opportunity to optimize their server selection beyond their current capabilities by communicating 
directly with the ISP (CP-ISP Communication, see Figure~\ref{fig:system-overview}).
Furthermore, ISPs gain the ability of adapting to the volatile traffic induced by content 
delivery, by being able to influence the choice of the CP. We believe that \cate is a 
step forward in improving the end-user performance and enabling ISP and CP collaboration.

\subsection{Incentives}\label{sec:Incentive}

The opportunities that \cate enables for both CPs and ISPs require that both parties have 
incentives to work together. Furthermore, the growing awareness of end-users about \cate's 
benefits will accelerate the penetration of \cate in a highly commoditized content delivery 
market.

\subsubsection{Incentives for CPs}\label{sec:CP-Incentive}

The market of CPs requires them to enable new applications while reducing their operational cost, 
and to improve the end-user experience~\cite{Akamai-Network}. With \cate improving the mapping
of end-users to servers, CPs can expect improvements in the end-user experience, and thus, a 
competitive advantage. This is particularly important for CPs in light of the commoditization 
of the content delivery market and the choice that is offered to end-users, for example through 
meta-CDNs~\cite{Conviva2011}. The improved mapping also yields better infrastructure amortization 
and thanks to \cate, CPs will no longer have to perform and analyze voluminous measurements in order 
to infer the network conditions or end-user locations.

To stimulate the use of \cate, ISPs can operate and provide \cate as a free service to CPs or even offer 
discounts on peering or hosting prices, \eg for early adopters and CPs that expose a higher server diversity 
while using \cate. The loss of peering or hosting revenue is amortized with the benefits of a lowered network 
utilization, reduced investments in network capacity expansion and by taking back some control over the 
traffic within the network. Ma et al.~\cite{CooperativeSettlement:ToN} have developed a methodology to 
estimate the prices in such a cooperative scheme by utilizing the Shapley settlement mechanism.
\cate can also act as an enabler for CPs and ISPs to jointly launch new applications in a cost-effective 
way, for example traffic-intensive applications such as the delivery of high definition video on-demand, 
or real-time applications such as online games. In an ISP-CP collaborative scheme, \cate can play the role 
of a recommendation system and is not intended to be applied unilaterally by the ISP.

\newpage
\subsubsection{Incentives for ISPs}\label{sec:ISP-Incentive}

ISPs are interested in reducing their operational and infrastructure upgrade costs, offering broadband 
services at competitive prices, and delivering the best end-user experience possible. Due to network 
congestion during the peak hour, ISPs in North America have recently revisited the flat pricing model 
and have announced data caps to broadband services. A better management of traffic in their network 
with \cate can allow them to offer higher data caps or even alleviate the need to introduce them. From 
an ISP perspective, \cate offers the possibility to do global traffic and peering management, through an 
improved awareness of the traffic across the whole network. For example, peering agreements with CPs 
can offer the use of \cate in exchange for reduced costs to the CPs. This can be an incentive for CPs to 
peer with a \cate-enable ISP and an additional revenue for an ISP, as such reduced prices can attract 
additional peering customers. An ISP can also offer \cate to other ISPs it peers with, which makes sense 
especially in the case that the peering ISPs hosts content or also acts as CP. The interaction and federation 
of CPs run by ISPs can also be enabled through \cate. There is high interest on the side of ISPs, as reflected 
by the creation of the IETF working group CDNi~\cite{ietf-cdni-protocol}. Furthermore, \cate has the 
potential to reduce the significant overhead due to the handling of customer complaints that often do not 
stem from the operation of the ISP but the operation of CPs~\cite{ControllingDataCloud}. With \cate, ISPs 
can identify and mitigate congestion, and react to short disturbances caused by an increased demand of 
content from CPs.

\subsubsection{Incentives for end-users}\label{sec:Users-Incentive}

\cate offers a way to empower end-users to obtain the best possible quality of experience. As such,
this creates an incentive for end-users to support the adoption of \cate by both ISPs and CPs.
For example, an ISP can offer more attractive products, \ie higher bandwidth or lower prices, since
it is able to better manage the traffic inside its network. Also, thanks to better traffic engineering,
ISPs can increase data caps on their broadband offers, making the ISP more attractive to end-users.
Moreover, CPs that utilize \cate can offer better quality of experience to end-users. This can be done 
through premium services based on \cate. For example, CPs delivering streaming services can offer 
higher quality videos to end-users thanks to better server assignment and network engineering. Also, 
applications running over the Internet can greatly benefit in their performance from 
\cate~(see Appendix~\ref{sec:Active-measurements}). This, in turn, gives end-users a good reason 
to choose \cate enabled services.

\begin{figure}[t]
  \center \includegraphics[width=.75\linewidth,angle=0]{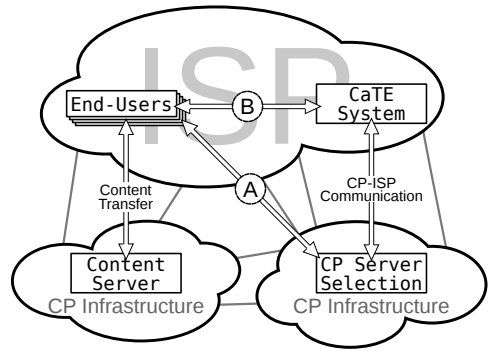}
  \caption{CaTE deployment and interaction with CPs.}
  \label{fig:system-overview}
  \vspace{-1.5em}
\end{figure}

\section{\textbf{\cate} Approach}\label{sec:Approach}

The concept of \cate relies on two key observations. First, a major fraction
of the traffic in ISPs is delivered by massively distributed CP infrastructures. Therefore, the same content
is often available at different network locations with different network paths to the end-user.
Second, the server selection of CPs is decoupled from the content transfer. 
Thus, it is possible to augment the server selection strategy of CPs with detailed information from ISPs 
about the current network state, the status of links that are traversed and the precise network location of the end-user.

\subsection{Concept of \textbf{\cate}}\label{sec:concept}

\cate relies on the fact that by selecting an appropriate server among those
being able to satisfy a request, the flow of traffic within the network can
be influenced. To illustrate the concept, we show in Figure~\ref{fig:flowSelection}
how, by selecting server A instead of B or C, a shorter path through the network is
chosen. However, CPs have limited knowledge about the path characteristics inside a network.
On the other hand, ISPs are aware of the state of their network, the location of their
users, as well as the path conditions between end-users and servers.
Given the large fraction of traffic that originates from CPs and their highly distributed
infrastructure, \cate can shift traffic among paths within a network and, through this,
achieve traffic engineering goals for both CPs and the ISP.

\subsection{\textbf{\cate} Deployment Schemes}\label{sec:deployment}

Our main architectural motivation is that the server selection is decoupled from the
content transfer. In Figure~\ref{fig:system-overview} we provide a simplified version
of how CPs handle content requests. Today, the server selection process of CPs
works as follows. When an end-user wants to obtain a specific content, it first sends a request to
the \textit{CP server selection} of the CP (see Figure~\ref{fig:system-overview}, \textit{(A)}). 
Today, there are two prevalent
techniques used to transfer this request: DNS queries and HTTP redirection.  The 
\textit{CP server selection} selects the content server based on the requested content, 
the objectives of the CP, its current view of the network, and its knowledge of the end-user 
network location. Finally, it returns the selected server IP, either through a DNS reply or a 
HTTP redirection, to the end-user, which in turn establishes a connection to the supplied server 
IP to download the content.

In order for \cate to hook into the server selection of CPs, a new component inside the
ISPs network is needed. In general, this component offers an interface between the CP and
the ISP to get supplement information about the network position of end-users, path conditions between
an end-user and eligible servers, etc. To this end, the system uses information readily
available to an ISP, such as the actual network topology, routing information,
end-user assignment databases, current network loads, etc. Today, systems capable of
providing the interface between an ISP and a CP are for example the IETF ALTO
service~\cite{ietf-alto-protocol} or the Provider-aided Distance information System (PaDIS)~\cite{PADIS2010}.
In Figure~\ref{fig:system-overview} we outline the range of possible \cate deployment schemes:

\noindent {\bf 1.~CP contacts ISP:~} The end-user contacts the CP server selection
    module via its DNS resolver \textit{(A)} as it does today. 
    When choosing the server for the end-user, the CP uses the \textit{CP-ISP Communication}
    to retrieve information about the network status, topology, or a recommendation
    by the ISP based on the network conditions between the end-user and the candidate
    content servers. The advantage of the recommendation option is that no party
    reveals any sensitive operational information.

    This can be implemented by including the client IP in
    the mapping request as proposed at the IETF dnsext
    working group~\cite{DNS-extension-IP-client} while using the IETF
    ALTO protocol or PaDIS by the CP to retrieve topology information,
    network status information, or server recommendation by the ISP.

\noindent {\bf 2.~ISP contacts CP:~} The end-user contacts \cate directly \textit{(B)}
    for the mapping. Then, \cate uses the \textit{CP-ISP Communication} to forward
    the request to the CP. The CP returns a list of potential servers and
    \cate ranks them based on network characteristics and the
    current path conditions between end-user and server network location.

    This can be implemented by utilizing the part of the DNS resolution
    process handled by CPs. When end-users query the ISP DNS resolver and, in turn,
    the CP DNS server, the CP returns all candidate content servers, which
    are re-ordered by the ISP DNS resolvers according to \cate.

\noindent {\bf 3.~ISP-based:~} The end-user contacts \cate directly \textit{(B)} for the mapping.
    However, \cate forwards the request through the \textit{CP-ISP Communication}
    to the CP server selection, which returns the normal reply as it happens
    today. \cate collects and aggregates the replies from the CP and overwrites
    the replies using the knowledge it has obtained from past results.

    This can be implemented by using the DNS resolution
    process of CPs. When end-users query the ISP DNS resolver the ISP
    forwards the request. However, the answer from the CP is kept and aggregated
    as proposed by Poese et al.~\cite{PADIS2010} and the DNS replies are
    overwritten as \cate sees fit.

\noindent {\bf 4.~User-based:~} The end-user collects the potential content servers from the
    CP as well as the current network state from the ISP. By utilizing this information,
    it calculates the best server to connect to based on active end-to-end measurements or
    previously reported experience. 

    This can be achieved when both the CP and the ISP
    run the IETF ALTO service or PaDIS. In this case, the client downloads all
    the needed information and performs the server selection itself.

\vspace{0.05in}
In the first three schemes \cate can be incrementally deployed and interacts with the
existing CP infrastructures while being transparent to the end-user.
In the collaborative schemes 1 and 2, the final decision is made by the CPs to avoid
any disturbance on their operation.
The frequency of ranking exchanges as well as the granularity of end-user location
identification is up to the administrator of the system.
It is also possible to provide end-users the choice to opt-in or opt-out.
CPs can also negotiate how many locations they make available to ISPs. Note, CPs can
dynamically change the locations made available to the ISP depending on the utilization of
each location. In the last deployment option, we describe how \cate can also be deployed at
the end-user, \eg via the browser or home gateway, but the penetration will be slower
as it requires the installation of software at the end-user.

\section{Modelling \textbf{\cate}}\label{sec:CaTE}

Next, we formalize \cate and discuss how it relates to traditional traffic engineering
and multipath routing.

\subsection{Traffic Engineering}\label{sec:Network-Model}

We model the network as a directed graph $G(V,E)$ where $V$ is the set of nodes
and $E$ is the set of links. An origin-destination (OD) flow $f_{od}$ consists of all traffic 
entering the network at a given point $o \in V$ (origin) and exiting the network at some 
point $d \in V$ (destination).
The traffic on a link is the superposition of all OD flows that traverse the link.

The relationship between link and OD flow traffic is expressed by the routing matrix $A$. The matrix
$A$ has size $|E|$ $\times$ $|V|^2$. Each element of matrix $A$ has a boolean value.
$A_{ml}=1$ if OD flow $m$ traverses link $l$, and $0$ otherwise. The routing matrix $A$ can be derived
from routing protocols, e.g., OSPF, ISIS, BGP. Typically, $A$ is very sparse since each OD flow traverses
only a very small number of links. Let {\bf y} be a vector of size $|E|$ with traffic counts on links and
{\bf x} a vector of size $|V|^2$ with traffic counts in OD flows, then {\bf y}=$A${\bf x}. Note, {\bf x}
is the vector representation of the traffic matrix.

\smallskip
\noindent{\bf Traditional Traffic Engineering:}\label{sec:TE-Definitions}
In its broadest sense, traffic engineering encompasses the application of
technology and scientific principles to the measurement, characterization,
modeling, and control of Internet traffic~\cite{Awduche_OverviewTE:2002}.
Traditionally, traffic engineering reduces to controlling and
optimizing the routing function and to steering traffic through the network
in the most effective way.
Translated into the above matrix form, traffic engineering is the process
of adjusting $A$, given the OD flows {\bf x}, so as to influence the link traffic {\bf y} in a desirable
way, as coined in~\cite{LakhinaSigmetrics2004}. The above definition assumes that the OD flow vector {\bf x}
is known. For instance, direct observations can be obtained, \eg with Netflow
data~\cite{Cisco_Netflow:99,TrafficDemand:ToN2001}.

\begin{figure}[t]
\center\includegraphics[width=0.90\linewidth]{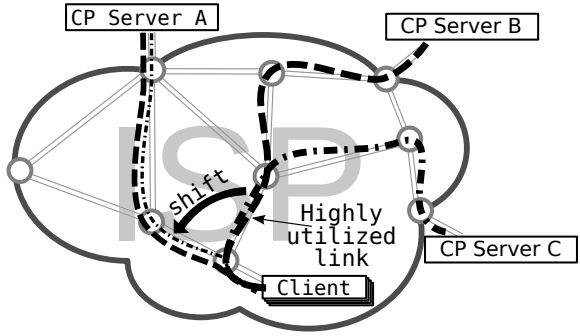}
\caption{Content-aware Traffic Engineering Process}
\label{fig:Content-aware-illustration2}
\vspace{-1.5em}
\end{figure}

\smallskip
\noindent {\bf Terminology:}\label{sec:Terminology}
We denote as {\it flow} an OD flow between two routers in the network. 
We call a flow {\it splittable} if arbitrarily small pieces of the flow can be assigned to other flows. 
This is not to be confused with end-to-end sessions, \ie TCP connections, which are \emph{un-splittable}. 
The assumption that flows are splittable is reasonable, as the percentage of traffic of a single
end-to-end session is small 
compared to that of a flow between routers. Let $C$ be the set of nominal capacities of the links in
the network $G$. We denote as {\it link utilization} the fraction of the link capacity that is used by 
flows. We denote as {\it flow utilization} the maximum link utilization among all links that a flow 
traverses. We introduce the terms of {\it traffic consumer} and {\it traffic producer} which refer to 
the aggregated demand of users attached to a router, and the CPs that are responsible for the traffic 
respectively. Throughout this paper, we refer to the different alternatives from which content can be 
supplied by a given CP as \emph{network locations} that host servers.

\subsection{Definition of \textbf{\cate}}\label{sec:CaTE-Definitions}

We revisit traffic engineering by focusing on the traffic demands rather than
changing the routing.

\noindent \textbf{Definition 1: Content-aware Traffic Engineering(\cate)} is the process of adjusting the traffic
demand vector {\bf x}, given a routing matrix $A$, so as to change the link traffic {\bf y}.

Not all the traffic can be adjusted arbitrarily. Only traffic for which location diversity is available can be adjusted by \cate.
Therefore, {\bf x}={\bf x$_r$}+{\bf x$_s$}
where {\bf x$_r$} denotes the content demands that can be adjusted and
{\bf x$_s$} denotes the content demands that can not be adjusted as there is only a single location in the network where
the content can be downloaded from.
The amount of traffic that can be adjusted depends on
the diversity of locations from which the content can be obtained.
We can rewrite the relation between traffic counts on links and traffic counts in flows as follows:
{\bf y}=$A$({\bf x$_s$} $+$ {\bf x$_r$}).
\cate adjusts the traffic on each link of the network by adjusting the content demands 
{\bf x$_r$}: {\bf y$_r$}=$A${\bf x$_r$}. Applying \cate means adjusting the content 
demand to satisfy a traffic engineering goal.

\noindent \textbf{Definition 2: Optimal Traffic Matrix} is the new traffic matrix,
{\bf x}$^*$, after applying \cate, given a network topology $G$, a routing matrix $A$ 
and an initial traffic matrix {\bf x}.

Figure~\ref{fig:Content-aware-illustration2} illustrates the \cate process.
A content consumer requests content that three different servers can
deliver. Let us assume that, without \cate, the CP redirects the clients to
servers B and C.
Unfortunately, the resulting traffic crosses a highly-utilized link. With \cate, 
content can also be downloaded from server A, thus, the traffic within the 
network is better balanced as the highly utilized link is circumvented.

Minimizing the maximum utilization across all links
in a network is a popular traffic engineering goal~\cite{FT00,FT01,ImprovingPerformanceInternet2009}.
It potentially improves the quality of experience and postpones the need for capacity increase. \cate mitigates
bottlenecks and minimizes the maximum link utilization by re-assigning parts of the traffic traversing
heavily loaded paths. Thus it redirects traffic to other, less utilized paths.
As we will elaborate in Section~\ref{sec:Evaluation}, different metrics such as path length or network delay
can also be used in \cate.

\subsection{\textbf{\cate} and Traditional TE}\label{sec:CaTE-TE}

\cate is complementary to routing-based traffic engineering as it does not
modify the routing. Routing-based traffic engineering adjusts routing
weights to adapt to traffic matrix changes. To avoid micro-loops during
IGP convergence~\cite{transient-IGP}, it is common practice to only adjust
a small number of routing weights~\cite{FT01}. To limit the number of changes
in routing weights, routing-based traffic engineering relies on traffic matrices
computed over long time periods and offline estimation of the routing weights.
Therefore, routing-based traffic engineering operates on time scales of hours,
which can be too slow to react to rapid change of traffic demands.
\cate complements routing-based traffic engineering and can
influence flows at shorter time scales by assigning clients to servers
on a per request basis. Thus, \cate influences the traffic
within a network online in a fine-grained fashion.

\subsection{\textbf{\cate} and Multipath Routing}\label{sec:CaTE-Multipath}

Multipath routing helps end-hosts to increase and control their upload 
capacity~\cite{PathSelection:CACM2011}. It can be used to minimize transit 
costs~\cite{Optimizing:Goldenberg2004}. Multipath also enables ASes to dynamically 
distribute the load inside networks in the presence of volatile and 
hard to predict traffic demand changes~\cite{TrafficDemand:ToN2001,MATE2001,TeXCP,Replex}.
This is a significant advantage, as routing-based traffic engineering can be too slow to 
react to phenomena such as flash crowds. Multipath takes advantage of the diversity of 
paths to better distribute traffic.

\cate also leverages the path diversity, and can be advantageously combined with multipath 
to further improve traffic engineering and end-user performance. One of the advantages of
\cate is its limited investments in hardware deployed within an ISP. It can be realized with 
no change to routers, contrary to some of the previous multipath proposals~\cite{TeXCP,MATE2001,Replex}.
The overhead of \cate is also limited as no state about individual TCP connections
needs to be maintained, contrary to multipath~\cite{TeXCP,MATE2001,Replex}.
In contrast to~\cite{MATE2001,TeXCP}, \cate is not restricted to MPLS-like solutions
and is easily deployable in todays networks.
\subsection{\textbf{\cate} and Oscillations}\label{sec:CaTE-Multipath}

Theoretical results~\cite{AdaptiveRouting2005,Convergence:Fischer2006} have shown that
load balancing algorithms can take advantage of multipath while provably avoiding traffic 
oscillations. In addition, their convergence is fast. Building on these theoretical results, Fischer 
et al. proposed REPLEX~\cite{Replex}, a dynamic traffic engineering algorithm that exploits 
the fact that there are multiple paths to a destination. It dynamically changes the traffic load 
routed on each path. Extensive simulations show that REPLEX leads to fast convergence, without 
oscillations, even when there is lag between consecutive updates about the state of the network.
\cate is derived from the same principles and thus inherits all the above-mentioned desired properties.

\section{\textbf{\cate} Algorithms}\label{sec:Algorithm}

In this section we propose algorithms to realize \cate, in the context of an ISP. A key observation is 
that \cate can be reduced to the restricted machine load balancing problem~\cite{RestrictedModel:1995}
for which optimal online algorithms are available. The benefit of the \cate online algorithm can be 
estimated either by reporting results from field tests within an ISP or by using trace-driven simulations. 
Typically, in operational networks only aggregated monitoring data is available. To estimate the 
benefit that \cate offers to an ISP, we present offline algorithms that uses traffic demands 
and server diversity over time extracted from those statistics as input.

\subsection{Connection to Restricted Machine Load Balancing}\label{sec:Reduction}

Given a set of CPs and their network location diversity, we consider the problem of re-assigning
the flows that correspond to demands of content consumers to the CPs in such a way that a specific 
traffic engineering goal is achieved. Given that sub-flows between end-systems and content pro\-vider 
servers can be re-distributed only to a subset of the network paths, we show that the solution of the 
optimal traffic matrix problem corresponds to solving the {\em restricted machine load balancing problem}~\cite{RestrictedModel:1995}.
In the restricted machine load balancing problem, a sequence of tasks is arriving, where each task 
can be executed by a subset of all the available machines. The goal is to assign each task upon arrival 
to one of the machines that can execute it so that the total load is minimized. Note, contrary to the case 
of multipath where paths between only one source-dest\-ination pair are utilized, \cate can utilize any 
eligible path between any candidate source and destination of traffic.

\begin{figure}[t]
\center\includegraphics[width=1\linewidth]{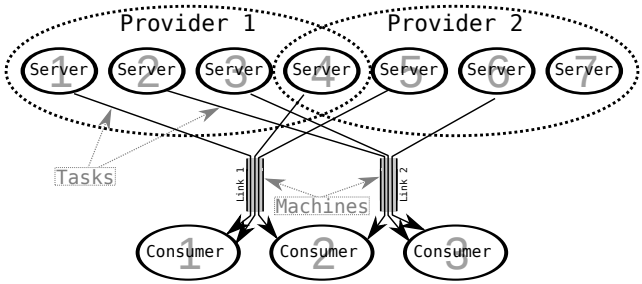}
\caption{\cate and Restricted Machine Load Balancing.}
\label{fig:theory-reduction}
\vspace{-1.5em}
\end{figure}

For ease of presentation let us assume that the traffic engineering goal is to minimize the maximum link 
utilization in the network~\cite{FT00,FT01}. Let us consider three consumers where each one wants to
download one unit of content from two different content providers, see Figure~\ref{fig:theory-reduction}.
Given that different servers can deliver the content on behalf of the two providers, the problem consists in 
assigning consumers to servers in such a way that their demands are satisfied while minimizing the maximum 
link utilization in the network. Thus, the problem is the restricted machine load balancing one where tasks 
are the demands satisfied by the servers and machines are the bottleneck links that are traversed when a path, 
out of all eligible server-consumer paths, is selected. Figure~\ref{fig:theory-reduction} shows one of the possible 
solutions to this problem, where consumer 1 is assigned to servers 1 and 4, consumer 2 to servers 5 and 2, and 
consumer 3 to servers 3 and 6. Note that the machine load refers to the utilization of the bottleneck 
links of eligible paths, denoted as link 1 and 2.

To be consistent with our terminology, we define the {\em restricted flow load balancing problem}.
Let $J$ be the set of the consumers in the network, $K$ be the set of content producers,
and $I$ be the set of servers for a given content provider, i.e., the set of locations where a request can be satisfied.
Note, this set is offered by the CP in order to satisfy its own objectives and can change over time.
We denote as $M_{jk}$ the set of flows that can deliver content for a given content producer $k$ to consumer $j$.

\noindent \textbf{Definition 3: Restricted Flow Load Balancing Problem}
is the problem of finding a feasible assignment of flows such that a traffic engineering goal is achieved,
given a set of sub-flows $\{f_{ijk}\}$ from all eligible servers $i \in I$ of a given content provider $k \in K$ 
to a consumer $j \in J$, and a set of eligible residual flows $f_{ij}^{-k},~i \in M_{jk}$ (after removing the 
traffic of the above mentioned sub-flows).

Despite some similarities, the nature of our problem differs from the multi-commodity flow and bin packing.
In the multi-commodity flow problem~\cite{MCFAL:1993}, the demand between source and destination pairs 
is given while in our problem the assignment of demands is part of the solution. In the bin packing 
problem~\cite{BinPacking:1997}, the objective is to minimize the number of bins, i.e., number of flows 
in our setting, even if this means deviating from the given traffic engineering goal. Note, in the restricted 
flow load balancing problem any eligible path from a candidate source to the destination can be used, contrary 
to the multipath problem where only equal-cost paths can be used.

\subsection{Online Algorithm and Competitiveness}\label{sec:Competitiveness}

We next turn to the design of online algorithms. It has been shown that in the online restricted machine 
load balancing problem, the greedy algorithm that schedules a permanent task to an eligible processor 
having the least load is exactly optimal~\cite{RestrictedModel:1995}, \ie it is the best that can be found, 
achieving a competitive ratio of $\lceil \log_2 n \rceil +1 $, where $n$ is the number of machines. If tasks 
are splittable then the greedy algorithm is 1-competitive, \ie it yields the same performance as an offline 
optimal algorithm. The greedy algorithm is an online one, thus it converges to the optimal solution immediately 
without oscillations.

In the restricted flow load balancing problem, the set $M_{jk}$ can be obtained from the set of candidate servers 
that can deliver content when utilizing \cate as described in Section~\ref{sec:deployment}. The online assignment 
of users to servers per request, which minimizes the overall load, leads to an optimal assignment of sessions within 
sub-flows. In our case, flows are splittable since the content corresponding to each content request is negligible 
compared to the overall traffic traversing a link. Note, the end-to-end TCP connections are not splittable. 
Thus, the following online algorithm is optimal:

\addtocounter{algocf}{1}
\noindent \textbf{Algorithm \arabic{algocf}. Online Greedy Server Selection.} Upon the arrival of a content user 
request, assign the user to the server that can deliver the content, out of all the servers offered by the CP, such that 
the traffic engineering goal is achieved.

\subsection{Estimating the Benefit of \textbf{\cate} with Passive Measurements}\label{sec:Gain-Network-Data}

Before applying \cate in real operational networks, it is important to understand the potential benefits that
it can bring in a given context. For example, the operator of an ISP network would like to know in advance what 
are the gains when applying \cate, as well as being able to answer what-if scenarios, when applying \cate to 
traffic delivered by different CPs. Operators of CPs would also like to quantify the benefits by participating 
in \cate before collaborating with an ISP. In most operational networks, aggregated statistics and passive 
measurements are collected to support operational decisions. Therefore, we provide a framework that allows 
a simulation-driven evaluation of \cate. To that end, we present offline algorithms that can take as input passive 
measurements and evaluate the potential gain when applying \cate in different scenarios 
in Appendix~\ref{sec:Algorithm-appendix}. We propose a linear programming formulation as well as greedy 
approximation algorithms to speed-up the process of estimating the gain when using \cate.

\section{Evaluation of \textbf{\cate}}\label{sec:Evaluation}

In this section, we quantify the potential of \cate with different traffic
engineering goals in mind. We evaluate \cate with operational data from
three different networks. For the first network, we rely on content demands
built from observed traffic of a European Tier-1 ISP. The other two networks,
namely AT\&T and Abilene, allow us to evaluate the impact of the ISP topology
structure.

\subsection{Experimental Setting}\label{sec:Experimental-Setting}

To evaluate \cate, an understanding of the studied ISP network is necessary,
including its topological properties and their implications on the flow of traffic.
Indeed, the topological properties of the ISP network influence the availability
of disjoint paths, which are key to benefit from the load-balancing ability of \cate.
Because \cate influences traffic aggregates inside the ISP network at the granularity
of requests directed to CPs, fine-grained traffic statistics are necessary.
Traffic counts per-OD flow, often used in the literature, are too coarse an input for \cate.

\begin{figure}[t]
\center\includegraphics[width=1\linewidth]{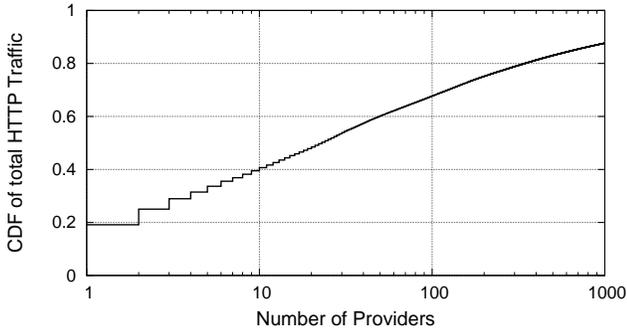}
\caption{CDF of traffic volume of CPs in ISP1.}
\label{fig:CDF-Traffic-CPs}
\vspace{-1.5em}
\end{figure}

\begin{figure}[!bpt]
\center\includegraphics[width=1\linewidth]{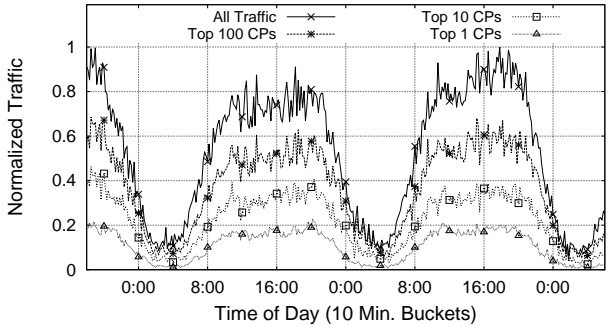}
\caption{Normalized traffic for top CPs by volume in ISP1.}
\label{fig:CPs-Total-Traffic}
\vspace{-1.5em}
\end{figure}

\subsubsection{Data from a Large European ISP}\label{sec:Residential-Traces}
To build fine-grained traffic demands, we rely on anonymized packet-level traces of
residential DSL connections from a large European Tier-1 ISP, henceforth called
\emph{ISP1}. For ISP1, we have the complete annotated router-level topology
including the router locations as well as all public and private peerings. ISP1 contains
more than $650$ routers and $30$ peering points all over the world.

We collect a $10$ days long trace starting on May 7, 2010. Our monitor, using Endace
monitoring cards~\cite{Endace}, allows us to observe the traffic of more than $20,000$
DSL lines to the Internet. We capture HTTP and DNS traffic using the Bro intrusion
detection system~\cite{paxson99bro}. We observe 720 million DNS messages as well
as more than 1 billion HTTP requests involving about 1.4 million unique hostnames,
representing more than 35 TBytes of data. With regards to the application mix, more
than 65\% of the traffic volume is due to HTTP. Other popular applications that contribute
to the overall traffic volume are NNTP, BitTorrent, and eDonkey.

A large fraction of the traffic in the Internet is due to large CPs, including CDNs, hyper-giants,
and OCHs, as reported in earlier studies~\cite{TrafficTypesGrowth:2011,arbor,PADIS2010}. In
Figure~\ref{fig:CDF-Traffic-CPs}, we plot the cumulative fraction of HTTP traffic volume as a
function of the CPs that originate the traffic. We define a CP as a organizational unit where
all servers from the distributed infrastructure serve the same content, such as Akamai or Google.
We rank the CPs by decreasing traffic volume
observed in our trace. Note that the x-axis uses a logarithmic scale. The top 10 CPs are responsible
for around 40\% of the HTTP traffic volume and the top 100 CPs for close to 70\% of the HTTP traffic
volume. The marginal increase of traffic is diminishing when increasing the number of CPs. This shows
that collaborating directly with a small number of large CPs, can yield significant savings.

In Figure~\ref{fig:CPs-Total-Traffic} we plot the traffic of the top 1, 10, 100 CPs by volume as
well as the total traffic over time normalized to the peak traffic in our dataset.
For illustrative purposes, we show the evolution across the first $60$ hours of our trace. A strong
diurnal pattern of traffic activity is observed. We again observe that a small number of CPs are
responsible for about half of the traffic. Similar observations are made for the rest of the trace.

\subsubsection{Understanding the Location Diversity of CPs}\label{sec:Traffic-by-Content-Provider}

To achieve traffic engineering goals, it is crucial to also understand the location
diversity of the top CPs, as \cate relies on the fact that the same content is available at
multiple locations.
Traffic originated from multiple network locations by a given CP is seen by \cate as a single atomic traffic
aggregate to be engineered. Furthermore, as routing in the Internet works per prefix, we assume that the
granularity of subnets is the finest at which \cate should engineer the traffic demand. Thus, we differentiate
candidate locations of CPs by their subnets and quantify the location diversity of CPs through the number
of subnets from which content can be obtained.

\begin{figure}[t]
\center\includegraphics[width=1\linewidth]{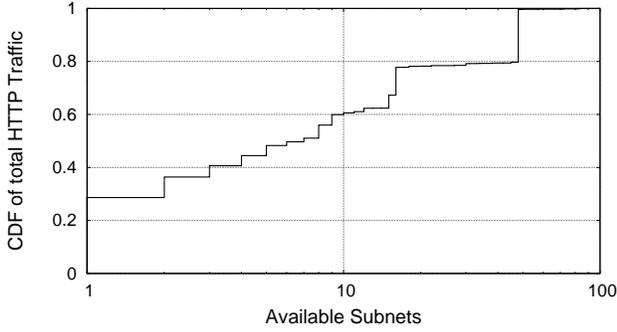}
\caption{Subnet diversity from which content is available.}
\label{fig:VolumeChoices}
\vspace{-1.5em}
\end{figure}

We examine the amount of location diversity offered by CPs based on traces from ISP1.
To identify the subnets of individual CPs, we rely on a similar methodology to the one from Poese
\etal~\cite{PADIS2010}. Our granularity is comparable to their "infrastructure redirection
aggregation". Figure~\ref{fig:VolumeChoices} shows the cumulative fraction of HTTP traffic
as a function of the number of subnets (logarithmic scale) from which a given content can be
obtained, over the entire $10$ days of the trace. We observe that more than $50\%$ of the
HTTP traffic can be delivered from at least $8$ different subnets, and more than $60\%$ of
the HTTP traffic from more than $3$ locations. These results confirm the observations made
in~\cite{PADIS2010}.

\subsubsection{Dynamics in Location Diversity}\label{sec:Server-Dynamics}

\begin{figure}[!bpt]
\center
\includegraphics[width=1\linewidth]{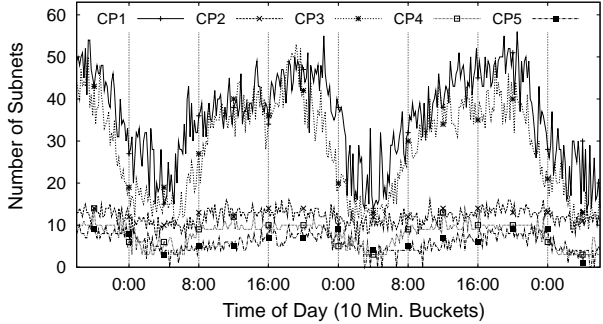}
\caption{Evolution over time of number of subnets for selected CPs in the top 10 CPs.}
\label{fig:TemporalEffect}
\vspace{-1.5em}
\end{figure}

So far the location diversity of CPs has been evaluated irrespective of time. To complement the
finding, we turn our attention to the location diversity exposed by CPs at small time-scales,
\ie in the order of minutes.
To this end, we split the original trace into $10$ minutes bins.  Figure~\ref{fig:TemporalEffect}
shows the evolution of the number of exposed subnets of five of the top 10 CPs by volume.
Note that the diversity exposed by some CPs exhibits explicit time of day patterns, while others
do not. This can be due to the structural setup or the type of content served by the CP.
The exposed location diversity patterns, \ie flat or diurnal, are representative for all CPs with a
major traffic share in our trace. We conclude that a significant location diversity is exposed by
popular CPs at any point in time, and is quite extensive during the peak hour.

\subsubsection{Content Demand Generation}\label{sec:TM}

The location diversity is not a mere observation about CPs deployment. It requires to revisit the
mapping between a given content demand and the realized traffic matrix. Given the location diversity for
content, multiple traffic matrices can be realized from a given content demand. The standard view of the
OD flows therefore provides an incomplete picture of the options available for \cate.

As an input for \cate, we introduce an abstraction of the demand that reflects the available location diversity.
We rely on the notion of \emph{potential vectors}, that were denoted as $x_{r}$ in Section~\ref{sec:CaTE-Definitions}.
To generate the potential vector for a given CP, the amount of traffic this CP originates as well as the potential
ingress points need to be known. Combining all potential vectors and $x_{s}$, we synthesize a network-wide
content demand matrix for each time bin, by scaling the traffic demand to match the network utilization of ISP1.
For our evaluation, we use the series of content demand matrices over a period of $10$ days. The content demands
are based exclusively on the HTTP traffic of our trace.

\subsection{CaTE in ISP1}\label{sec:CaTE-Main-Results}

\begin{figure}[t]
  \center \includegraphics[width=1\linewidth,angle=0]{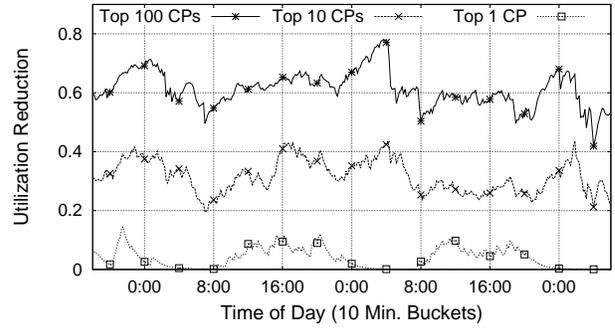}
  \center \includegraphics[width=1\linewidth,angle=0]{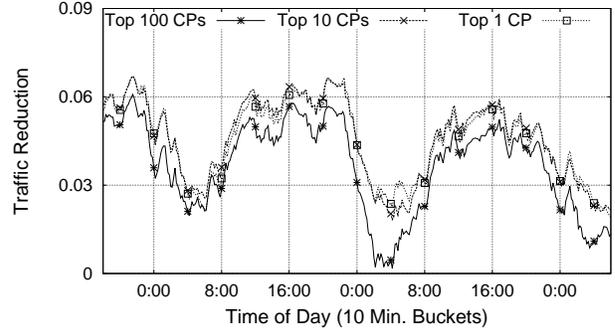}
  \caption{Maximum link utilization reduction (top) and total traffic reduction (bottom) with \cate for the top CPs.} 
  \label{fig:DT-Reduction_top1-100}
\vspace{-1.5em}
\end{figure}

To quantify the benefits of \cate, we first consider one of the most popular traffic engineering goals,
namely minimizing the maximum utilization of the links in the network~\cite{FT00,FT01}. The rationale
is that by minimizing the maximum link utilization, network bottlenecks are reduced, in turn limiting queueing
delays, improving the quality of experience and postponing the need for increased network capacity.

With \cate, an ISP can collaborate with any CP. It is up to the ISP to select the set of CPs that are the most
important to establish collaboration with. Since a significant fraction of the traffic originates from a small number
of CPs, we consider the most popular CPs by volume to evaluate \cate. In the following, we perform a sensitivity
study where we quantify the benefits of \cate when restricting its use to the top 1, 10 and 100 CPs by volume.
All other traffic remains unaffected by \cate. For all experiments, we use the Algorithm~\ref{alg:Greedy-Sort-Flow}
from Appendix~\ref{sec:Greedy}.

\noindent{\textbf{Effect on Maximum Link Utilization}}.
Figure~\ref{fig:DT-Reduction_top1-100} (top) shows the reduction of the maximum link utilization
over a period of $2$ days when considering the top 1, 10 and 100 CPs. Once again,
we normalized the absolute link utilization by the maximal one. The largest gain in maximum
link utilization reduction is up to $15\%$, $40\%$ and $70\%$ respectively. We observe large fluctuations
of the gains which are due to variations in traffic (see Figure~\ref{fig:VolumeChoices}) and location
diversity (Figure~\ref{fig:TemporalEffect}) throughout the day. The largest gains are obtained during
peak times, when there is more traffic and the highest location diversity is available. This is also when
congestion is at its peak and \cate is most needed. Our results show that \cate is able to react to diurnal
changes in traffic volume and utilizes the available location diversity.

\noindent{\textbf{Effect on Network-wide Traffic}}.
Although optimizing for link utilization, \cate reduces the overall traffic that flows through the network,
see Figure~\ref{fig:DT-Reduction_top1-100}~(bottom).
This is due to \cate choosing the shortest path when multiple ones with the same utilization are available,
thus, as a side effect, content is fetched from closer locations and therefore traverses less links.
With \cate, the gains in overall traffic reduction are up to $6\%$ and follows a clear diurnal pattern.
It is worth noticing that just with the top 10 CPs, the total traffic reduction is very close to the
one when considering the top 100 CPs, indicating that \cate only needs to be implemented with the major players.
Also, an ISP that is able to reduce the overall traffic inside its network is more competitive as it can serve
more end-users with the same infrastructure, delay additional investments in capacity upgrades and improve
end-user satisfaction.

\begin{figure}[t]
\center\includegraphics[width=1\linewidth]{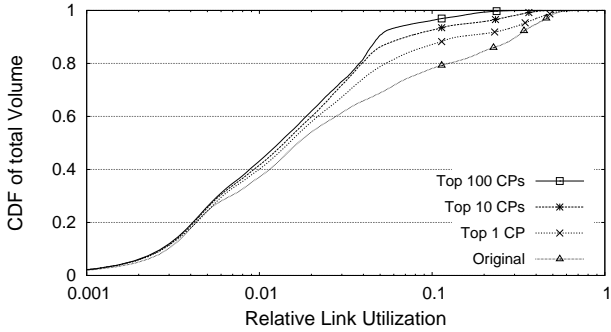}
\caption{Improvements in link utilization with \cate.}
\label{fig:DT-utilization-top1-100}
\vspace{-1.5em}
\end{figure}

\noindent{\textbf{Effect on Distribution of Link Utilization}}.
Reducing the maximum link utilization shifts traffic away from congested links.
However, it should not be done at the expense of creating congestion on other highly utilized links.
In Figure~\ref{fig:DT-utilization-top1-100} we plot the CDF of traffic volume in ISP1 across all link utilizations,
normalized by the maximum one when considering sets of the top CPs by volume.
The results show that \cate shifts the traffic away from highly utilized links to
low utilized ones.

\begin{figure}[t]
 \center\includegraphics[width=1\linewidth]{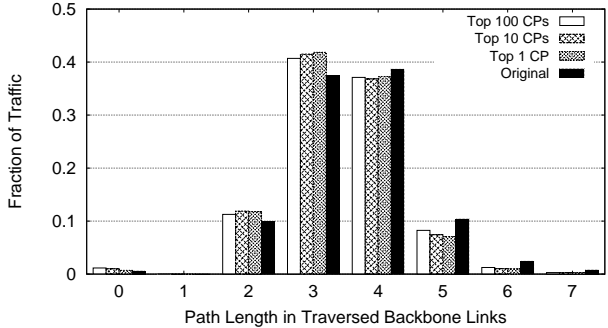}
 \caption{Backbone path length count with \cate.} 
 \label{fig:CDF-hopcount-top1-100}
 \vspace{-1.5em}
\end{figure}

\noindent{\textbf{Effect on Traffic Path Length}}.
Our results in Figure~\ref{fig:DT-Reduction_top1-100} (bottom) show a reduction in the overall traffic in ISP1,
which can be attributed to an overall reduction of the path length. Path length reduction is an important metric
for ISPs for the dimensioning of the network as well as the reduction of operational costs. To quantify this
reduction in terms of the path length inside ISP1, Figure~\ref{fig:CDF-hopcount-top1-100} shows the relative traffic
across different path lengths inside the network.
\cate redirects the traffic towards paths with the same or even shorter length than the ones used without \cate,
only in the rare case where a longer paths yields a lower utilization, \cate can choose a longer one.
Note that there is no traffic for backbone path length equal to 1 due to the network design of ISP1.
We conclude that applying \cate to a small number of CPs yields major improvements in terms of path length.

\begin{figure}[t]
  \center \includegraphics[width=1\linewidth,angle=0]{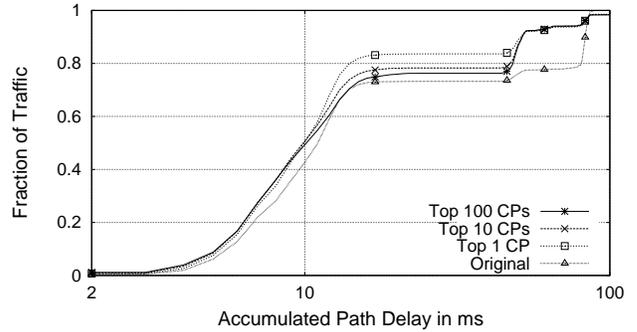}
  \caption{Improvement in path delay (in ms) with \cate.} 
  \label{fig:DT-Delay_top1-100}
\vspace{-1.5em}
\end{figure}

\noindent{\textbf{Effect on Path Delay}}. Although the objective of minimizing maximum link utilization is not
directly related to the reduction of path delay, the achieved reduction in path length directly affects the path delay.
Figure~\ref{fig:DT-Delay_top1-100} shows the accumulated path delay for the traffic that flows
within ISP1, when applying \cate. The reported numbers
for the backbone path delay are relatively modest compared to the values for the access part of
the network~\cite{OnDominantCharacteristics2009}. However, improving the access delay requires significant
investments as it can be done mostly through changes in the access technology, \eg from copper to fiber.
When considering the end-to-end delay, the delay of the path outside the ISP's network also needs
to be considered. As content infrastructures are located close to peering
points~\cite{arbor,MovingBeyondE2E2009,Cartography}, \eg IXPs or private peerings, the delays are expected
to be relatively small, especially for popular CPs. Estimating the impact of \cate on the end-to-end performance
for every application is very challenging, due to the many factors that influence flow performance, especially
network bottlenecks outside the considered ISP. In Appendix~\ref{sec:Active-measurements}
we show the results from active measurements conducted in the case of traffic-heavy applications, confirming
the significant improvements in end-to-end delay as well as download time that can be achieved thanks to \cate.

\noindent{\textbf{Summary}}. Our evaluation shows that \cate yields encouraging results, even when only a
few large CPs are collaborating with an ISP. In fact, even metrics that are not directly related to the optimization
function of \cate are improved. Besides significant improvements for the operation
of ISP networks, the end-users are expected to also benefit from these gains. This can be attributed to the decrease
of delay as well as the reduced link utilization.

\subsection{\textbf{\cate} with other Network Metrics}\label{sec:other-metrics}

So far we have evaluated \cate with one traffic engineering objective, namely, the minimization of
maximum link utilization. \cate allows ISPs and CPs to
to optimize for other network metrics such as path length or path delay.
To this end, we quantify the effects of \cate when using path length and delay and compare it with
the results presented in Section~\ref{sec:CaTE-Main-Results}.
We focus on the top 10 CPs as our results show that most of the benefits from \cate can be achieved with this
rather low number of CPs. Similar observations are made when applying \cate to the top 1 and 100 CPs.

\begin{figure}
  \center \includegraphics[width=1\linewidth,angle=0]{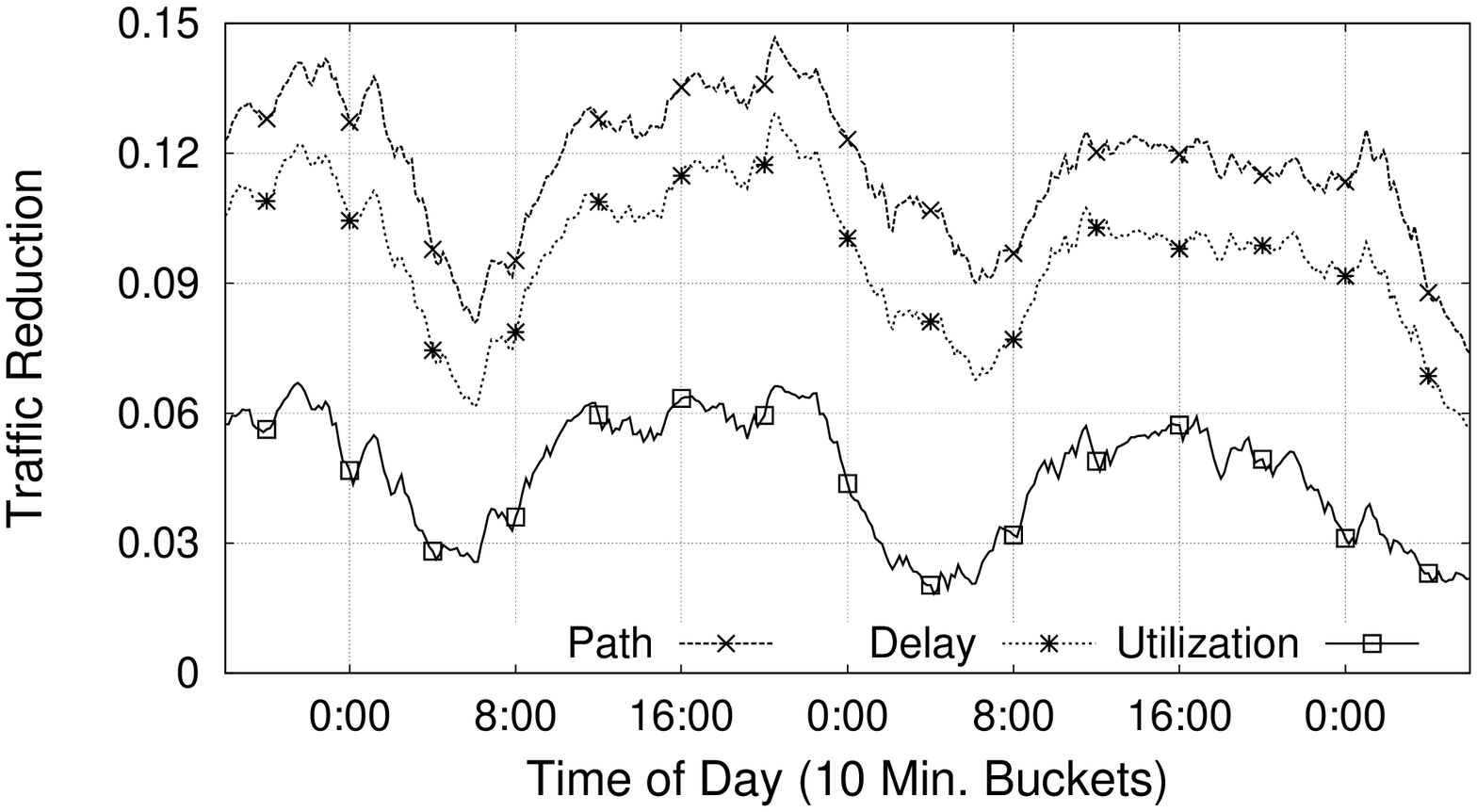}
  \center \includegraphics[width=1\linewidth,angle=0]{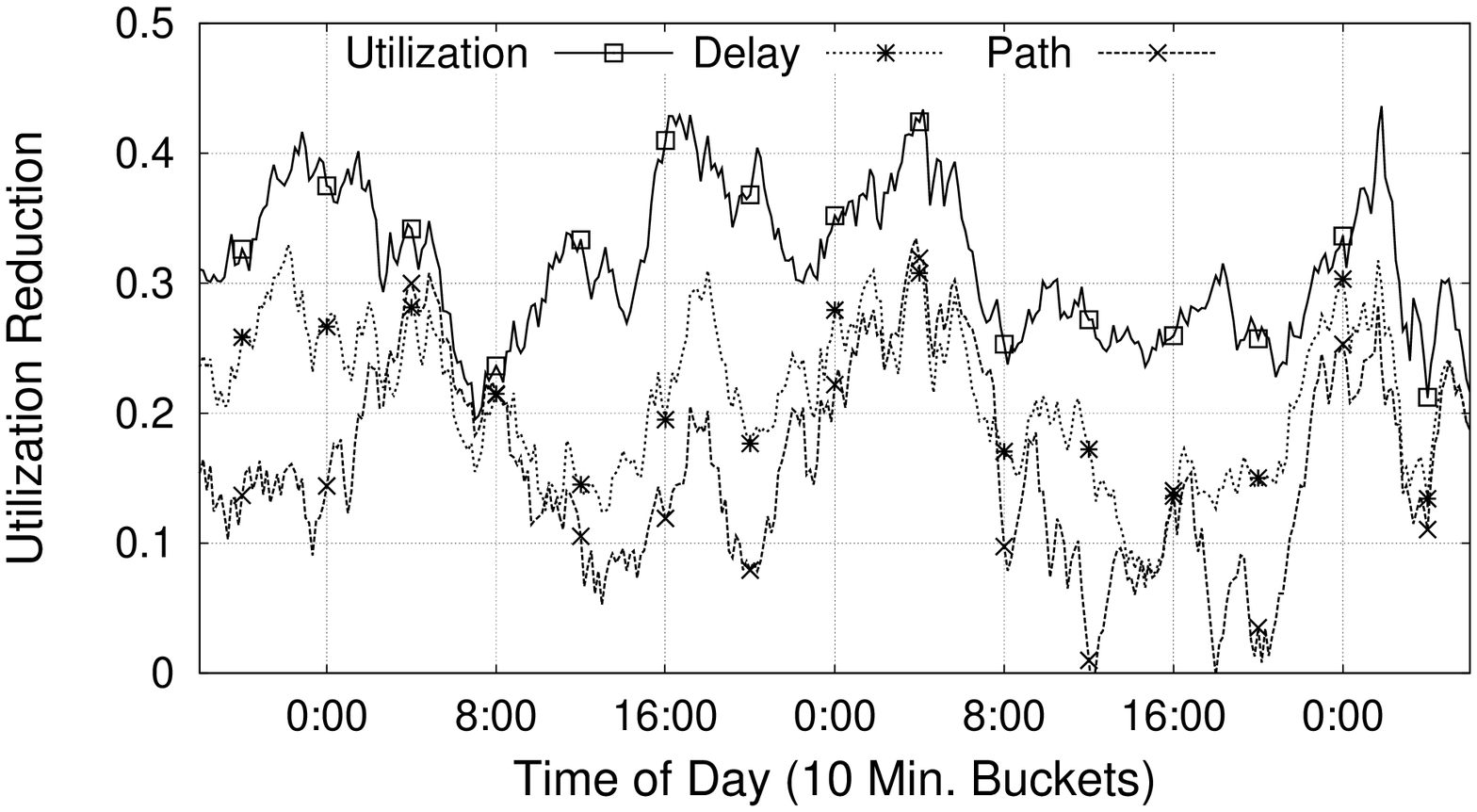}
  \caption{Total traffic (top) and maximum link utilization (bottom) reduction with \cate and different network metrics.}
  \label{fig:traffic-reduction-link-reduction-top10-all-metrics}
\vspace{-1.5em}
\end{figure}

\begin{figure}[t]
  \center \includegraphics[width=1\linewidth,angle=0]{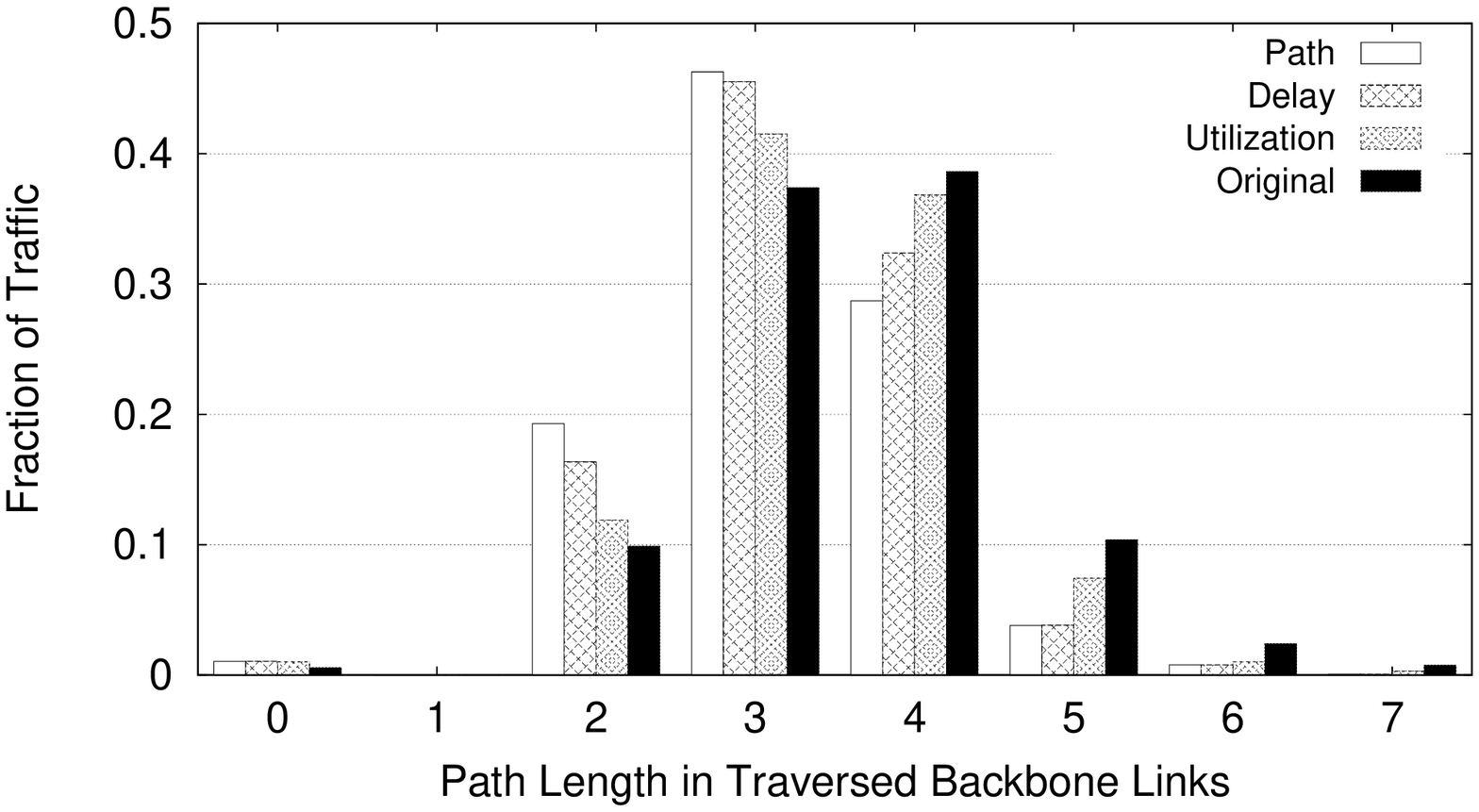}
  \center \includegraphics[width=1\linewidth,angle=0]{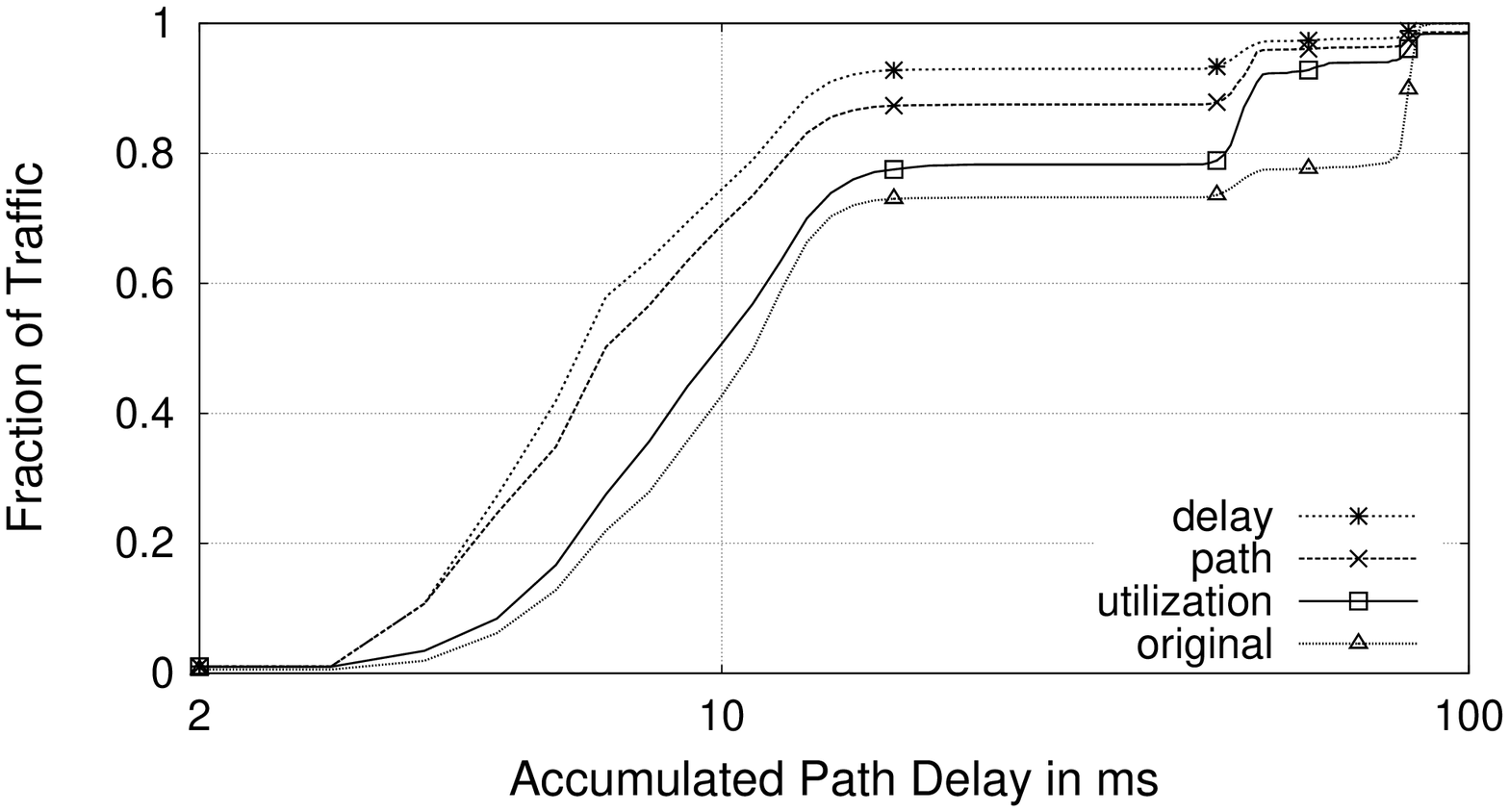}
  \caption{Backbone path length (top) and accumulated path delay (bottom) with \cate and different network metrics.}
  \label{fig:path-length-delay-top10-all-metrics}
\vspace{-1.5em}
\end{figure}

In Figure~\ref{fig:traffic-reduction-link-reduction-top10-all-metrics}~(top) we plot the total traffic reduction when applying
\cate to the top 10 CPs with different optimization goals. The first observation is that when the network
metric is path length, the total traffic reduction is the highest, with up to $15$\%. The total traffic reduction when optimizing
for path length are close to the one achieved when the metric is delay. Optimizing for other metrics provides
the expected result: the optimized metric is significantly improved, but at the cost of not optimizing other metrics as much.
For example, optimizing for link utilization diminishes the benefits from path length
(Figure~\ref{fig:path-length-delay-top10-all-metrics}~top) and vice-versa
(Figure~\ref{fig:traffic-reduction-link-reduction-top10-all-metrics}~bottom).
Still, significant improvements can be achieved even when optimizing for another network metric and we encountered no case of
significant deterioration in on of the network metrics throughout our experiments, see Figure~\ref{fig:traffic-reduction-link-reduction-top10-all-metrics}
and Figure~\ref{fig:path-length-delay-top10-all-metrics}.

\subsection{\textbf{\cate} in AT\&T and Abilene}\label{sec:Cate-in-other-ISPs}

To quantify the potential benefits of \cate in networks with different topological
structures than ISP1, we repeat our experiments for two other ISPs: AT\&T and Abilene.

{\textbf{AT\&T}} is one of the largest commercial networks.
We use the topology for the US backbone of AT\&T 
as measured by the Rocketfuel project~\cite{Rocketfuel:ToN2004,DisCarte:Sigcomm2008}.
Given that no publicly available traffic demands exist for AT\&T, we rely on the gravity
model~\cite{roughan-synthesis} to generate several traffic demand matrices as in ISP1.

{\textbf{Abilene}} is the academic network in the US.
We use the Abilene topology and traffic demands covering a 6 month period that are both publicly
available.\footnote{\small \url{http://userweb.cs.utexas.edu/~yzhang/research/AbileneTM/}}

The topology of both networks differ significantly from the one of ISP1.
In AT\&T, many smaller nodes within a geographical area are aggregated into a larger one.
Abilene has few but large and well connected nodes with a high degree of peerings.
For the application mix we rely on recent measurements in AT\&T~\cite{TrafficTypesGrowth:2011}
and
for server diversity we rely on measurements of users in these networks~\cite{Cartography}.

\begin{figure*}[!tpb]
\center\includegraphics[width=0.48\linewidth]{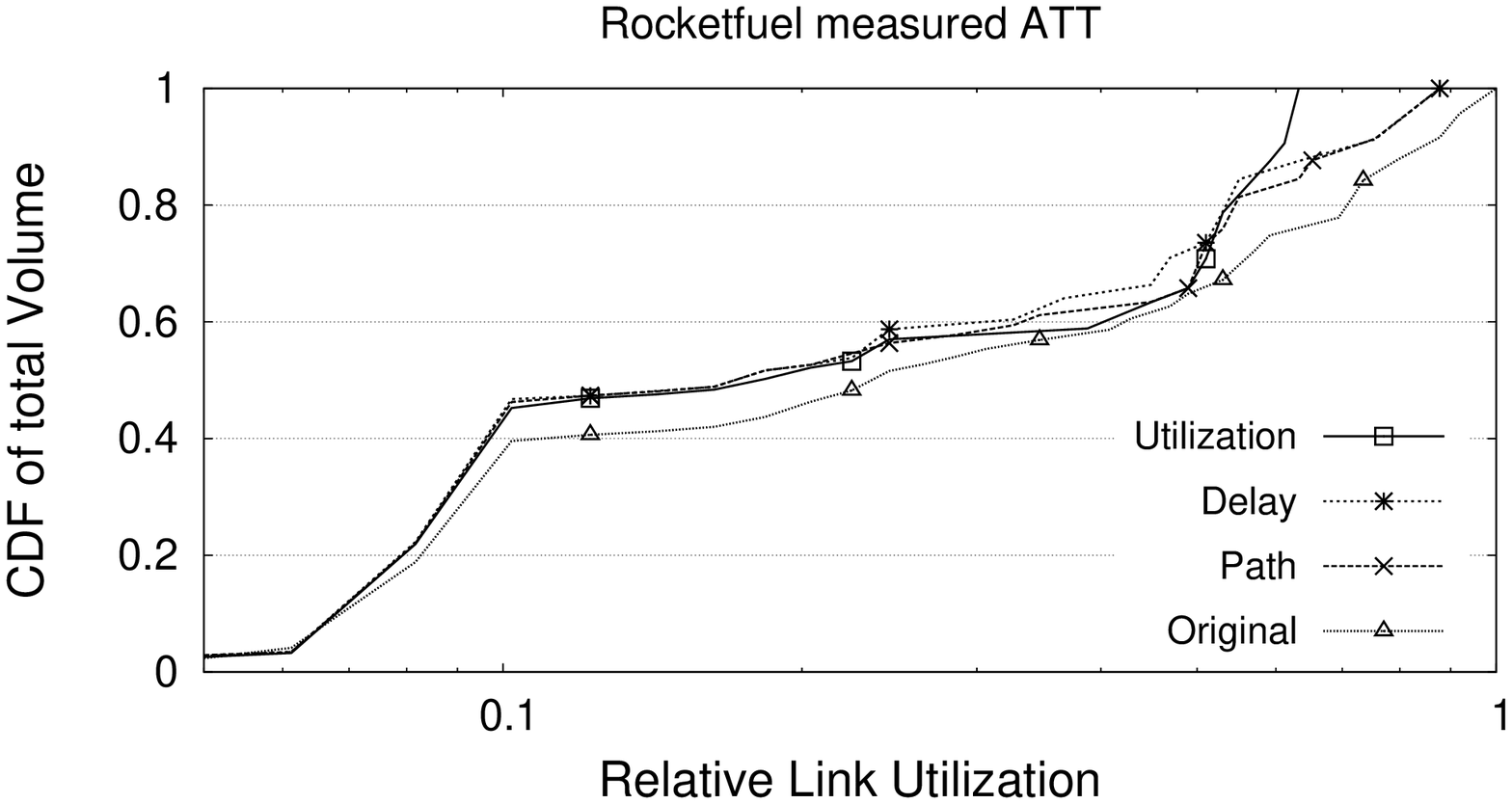}
\includegraphics[width=0.48\linewidth]{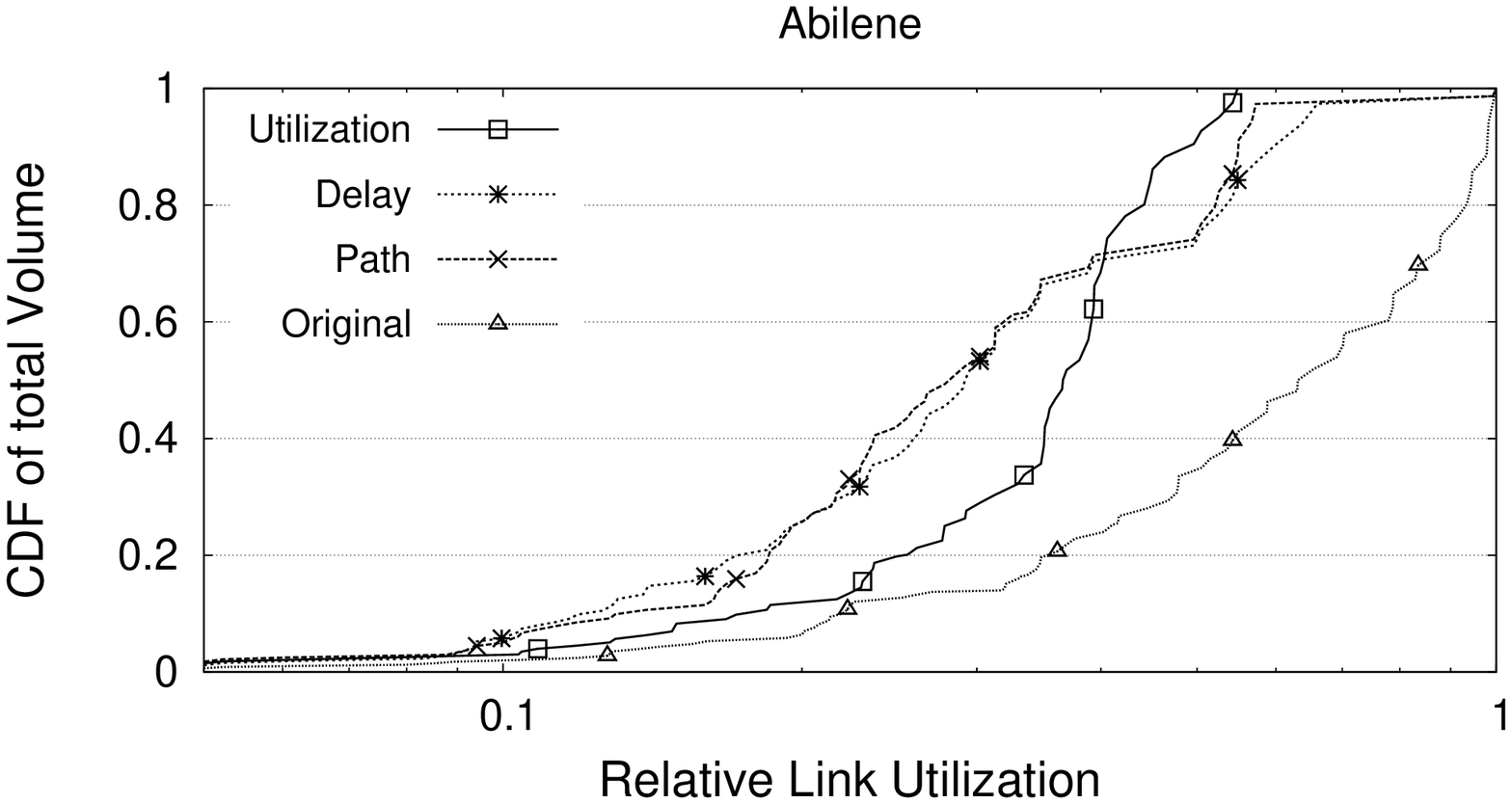}
\caption{Link utilization improvements after applying when using \cate in AT\&T and Abilene.}
\label{fig:ATT-Abilene}
\vspace{-1.5em}
\end{figure*}

Figure~\ref{fig:ATT-Abilene} shows the cumulative fraction of normalized link utilizations for AT\&T
and Abilene with different optimization goals. As already done in ISP, only the Top 10 CPs are
considered for \cate, while all other traffic stays unaffected.
For AT\&T the benefit for the maximum link utilization is about $36\%$ when the network
is optimized for minimizing the maximum link utilization, while the median
reduction in terms of network-wide traffic is about $3.7\%$. When other optimizations are used,
the benefits of \cate regarding the link utilization minimization are approximately $12\%$
for path length and delay. However, when looking at the median traffic reduction of these metrics,
the traffic is reduced by $5.4\%$ when path length is used, while delay achieves a reduction of $5\%$.
In the Abilene network benefits of \cate are more significant: $45\%$ reduction in the maximum link
utilization and $18\%$ for network-wide traffic when \cate optimizes for link utilization. 
When targeting
the other two metrics, \ie path length and delay, the results show that \cate does not reduce the maximum link
utilization. In fact, the maximum link utilizations stays constant. This is due to the structure of the network
and the fact that the content is available closer, but at the cost of keeping the high utilization on some of the links.
However, when looking at the median traffic reduction, both metrics manage to reduce the traffic
by over $24\%$. These results show that \cate is capable of targeting different optimization goals
in different network structures and is able to optimize for different metrics.

It is worth noting that for AT\&T $40\%$ of the links have a normalized link utilization less than $10\%$ while
the remaining link utilizations are distributed almost linear. This distribution fits the structural observations
made for the AT\&T network: many links from smaller nodes are aggregated into larger ones. This also explains why
the benefits for AT\&T are smaller, since such a structure reduces the path diversity. Turning our attention to Abilene,
we attribute the higher reduction of maximum link utilization and network-wide traffic to the non-hierarchical
structure of the network and a higher ratio of peering locations. Applying \cate to both AT\&T and Abilene networks where
the network metric is delay or path length shows similar behavior of \cate as it does in ISP1.

\subsection{\textbf{\cate} and Popular Applications}\label{sec:Discussion}

Today, the launch of new content hosted on CPs such as high definition video or others that share
flash-crowd characteristics, is not done in coordination with ISPs. 
This is challenging to ISPs that have to deal with rapid shifts of traffic volume as currently deployed traffic
engineering tools are too slow to react to rapid demand changes. Furthermore, the end-user experience for
popular applications is far from optimal as application designers have limited means to optimize the end-to-end delivery 
of content~\cite{MovingBeyondE2E2009}. Both ISPs and applications would benefit from the improved traffic
engineering capabilities of \cate. We believe that \cate can act as an enabler for ISP-application collaboration.

For example, Netflix, a very popular application that delivers high quality videos to end-users, relies on commercial CDNs 
such as Level3 and Limelight to improve the content delivery. Today, Netflix is only available in North and Latin America.
However, Netflix has announced that it will be launching its services in Europe early 2012. 
To quantify the effect of Netflix coming to Europe, we use our simulation to estimate the effect on ISP1. 
We run a series of experiments, assuming that the traffic of the CPs hosting Netflix will increase 20-fold. 
Our results show that with \cate, the total HTTP traffic volume is reduced by up to 8\% and 
the utilization of the most utilized link by 60\%.
More detailed results can be found in Appendix~\ref{sec:Netflix}.
\section{Related Work}\label{sec:Related-Work}

To meet the requirements of mission critical applications with stringent Service 
Level Agreements (SLAs), today's ISPs rely on traffic engineering~\cite{Awduche_OverviewTE:2002} 
to better control the flow of IP packets inside their network. Several techniques have
been proposed in the literature, some require tuning of the IP routing protocols
used inside the ISP network~\cite{FT00,FT01,Wang_Internet:2001}, while others 
rely on multipath~\cite{TrafficDemand:ToN2001,MATE2001,TeXCP,Replex,Optimizing:Goldenberg2004,GreenTE}.
Changing routing weights can lead to oscillations~\cite{griffin02sigcomm} and
is applied on time scales of hours. Multipath enables ISPs to dynamically distribute 
the traffic load within the network in the presence of volatile and hard to predict 
traffic demand changes~\cite{TrafficDemand:ToN2001,MATE2001,TeXCP,Replex},
even at very small time scales, but requires additional configuration and management or 
router support. \cate is complementary to both routing-based traffic engineering and multipath
enabled networks.

Traffic engineering relies on the availability of information about the traffic demands, 
which can be obtained either by direct observations~\cite{TrafficDemand:ToN2001,gunnar_estimation:2004,inferring-web-demand,uhlig-public} 
or through inference~\cite{medina_tm:2002,zhang_link:2003,zhang_tm:2003,soule_large_elements:2004,independent-TM}.
\cate relies on the network location diversity exposed by current hosting and content delivery 
infrastructures~\cite{Cartography}.

Game-theoretic results~\cite{CooperativeISPCDN,TECDN,CooperativeSettlement:ToN} show 
that the collaboration between CPs and ISPs can lead to a win-win situation. Recent studies also 
show that content location diversity has significant implications on traffic engineering within 
an ISP~\cite{BeyondMLU}. To our knowledge, \cate is the first system that is proposed to leverage 
the benefits of a direct collaboration between CPs and ISPs. 

\section{Summary}\label{sec:Conclusion}

Today, a large fraction of Internet traffic is due to a few content providers that rely on 
highly distributed infrastructures~\cite{arbor,ImprovingPerformanceInternet2009,Cartography}.
These distributed infrastructures expose a significant location diversity, which opens 
new opportunities to improve end-user performance, help CPs to better locale end-user 
and circumvent network bottlenecks, and enables new traffic engineering capabilities. 
We introduce the concept of \textit{content-aware traffic engineering} (\cate), that 
leverages this location diversity to engineer the traffic through careful selection of the 
locations from which content is obtained. We propose deployment schemes of \cate based 
on an online algorithm. The algorithm is stable and incurs no oscillations in link 
utilizations. Furthermore, \cate works on time scales ranging between the TCP control 
loop and traditional traffic engineering, and therefore advantageously complements 
existing traffic engineering techniques.

We evaluate some of the potential benefits of \cate on multiple operational networks
using an offline derivative of the online algorithm. Our results show that \cate 
provides benefits to CPs, ISPs and end-users, by reducing the maximum link utilization, 
the path length and the delay inside an ISP network, as well as enabling improved 
end-user to CP server assignment.

In the future, we envision \cate as an enabler for coordinated and Internet-wide 
traffic engineering. Meanwhile, \cate creates incentives for both ISPs and CPs to 
interlock their traffic engineering planes through the mutual benefits it brings. As 
further work, we want to deploy a prototype implementation of \cate and evaluate it 
through a direct collaboration between a CP and an ISP.
{\small
\bibliographystyle{plain}
\bibliography{paper}
}

\newpage
\begin{appendix}
\section{Estimating the Benefits of {\bf \cate}\\ with Passive Measurements}\label{sec:Algorithm-appendix}

We answer the question of the potential benefit \cate can offer to CPs, ISPs, and end-users.
The online algorithm requires deployment of \cate inside an operational network.
An alternative is to rely on a simulation-driven evaluation of \cate. For this, we
design offline algorithms that take as input passive measurements and estimate the gain
when applying \cate under different scenarios. We first propose a linear programming formulation
and then we present greedy algorithms to speed-up the process of estimating the benefits of \cate.

\subsection{Linear Programming Formulation}\label{sec:LP}

To estimate the potential improvement of \cate we formulate the Restricted Flow Load Balancing
problem (see Section~\ref{sec:Reduction}) as a Linear Program (LP) with restrictions on the variable values.
Variables $f_{ijk}$ correspond to flows that can be influenced. Setting $f_{ijk}=0$
indicates that consumer $j$ cannot download the content from server $i$ of a content provider $k$.
For each consumer $j$ we require that its demand $d_{jk}$ for content provider $k$ is satisfied,
\ie we require $\sum_{i\in M_{jk}} f_{ijk} = d_{jk}$.
The utilization on a flow $f_{ij}$ is expressed as $f_{ij} = \sum_k f_{ijk} $.

We use the objective function to encode the traffic engineering goal.
For ease of presentation we use as objective function the minimization of the maximum link utilization.
Let $T_e$ be the set of flows $f_{ij}$ that traverse a link $e \in E$.
The link utilization of a link $e \in E$ is expressed as $L_e = \sum_{T_e} f_{ij}$.
Let variable $L$ correspond to the maximum link utilization.
We use the inequality $\sum_{T_e} f_{ij} \leq L$ for all links. 
This results in the following LP problem:

\begin{align*}
  &{\tt min}~L            &&  \\
  &\sum_i f_{ijk}=d_{jk},   && \forall~j\in J,~k\in K\\
  &\sum_{T_e} f_{ijk}\leq L,    && \forall~j\in J,~i\in I,~k \in K,~e\in E\\
  &0\leq f_{ijk} \leq d_{jk}, && \forall~j\in J,~i\in M_{jk},~k \in K\\
  &f_{ijk}=0,       && \forall~j\in J,~i\notin M_{jk},~k \in K
\end{align*}\\

The solution of the above LP provides a fractional assignment of flows under the assumption
that flows are splittable and thus can be solved in polynomial time~\cite{KLinearProgramming1979}.
The solution is the optimal flow assignment, $f_{ijk}^*$, that corresponds to the optimal traffic matrix {\bf x}$^*$.
If flows are not splittable, or the sub-flows are discretized,
then the integer programming formulation has to be solved.
In this case the Restricted Flow Load Balancing problem is
NP-hard and a polynomial time rounding algorithm that approximates the assignment within a factor of $2$
exists~\cite{LSTApproximation:MP1990}.

\subsection{Approximation Algorithms}\label{sec:Greedy}

Since it is a common practice for operators to study multiple scenarios to
quantify the effect of changes in traffic matrices over periods
that spans multiple weeks or months, solutions based on LP may be too slow.
It might be also too slow to estimate the gain of \cate when applying it to an arbitrary combination of CPs.
To that end, we turn our attention to the design of fast approximation algorithms.
Simple greedy algorithms for load balancing problems~\cite{Graham:1969} are among the best known.
Accordingly, we propose a greedy algorithm for our problem which starts with
the largest flow first.

\begin{algorithm}[t]
\footnotesize \caption{\bf Iterative Greedy-Sort-Flow.}
\begin{algorithmic}[]
\State\hspace{-0.32in}{{\bf INPUT: $I$, $J$, $K$, $\{f_{ijk}\}$, $\{M_{jk}\}$, $A$}.}
\State\hspace{-0.32in}{\bf OUTPUT: $\{f_{ijk}^*\}$.}
\vspace{0.1in}
\State\hspace{-0.32in}{\bf Initialization:}
\State\hspace{-0.32in}{1.~Sort $k \in K$ by decreasing volume: $\sum_i \sum_j f_{ijk}$.}
\State\hspace{-0.32in}{2.~Sort $j \in J$ by decreasing volume: $\sum_i f_{ijk}$ for all $k \in K$.}
\vspace{0.1in}
\State\hspace{-0.32in}{\bf Iteration:}
\State\hspace{-0.32in}{Until no sub-flow is re-assigned or the maximum number of\\\hspace{-0.32in}iterations has been reached.}
\State\hspace{-0.32in}{~~~$\rhd$ Pick unprocessed $k \in K$ in descending order.}
\State\hspace{-0.32in}{~~~~~~~$\rhd$ Pick unprocessed $j \in J$ in descending order.}
\State\hspace{-0.32in}{~~~~~~~~~~$\rhd$ Re-assign $f_{ijk}$ in $f_{ij}^{-k},~i \in M_{jk}$~s.t. the engineering\\
~~~~goal is achieved.}
\end{algorithmic}
\label{alg:Greedy-Sort-Flow}
\end{algorithm}

\addtocounter{algocf}{1}
\noindent \textbf{Algorithm \arabic{algocf}: Greedy-Sort-Flow.} Sort sub-flows in decreasing order based on volume
and re-assign them in this order to any other eligible flow which, after assigning the sub-flow $f_{ijk}$, will yield
the desire traffic engineering goal.

Assignment in sorted order has been shown to significantly improve the
approximation ratio and the convergence speed~\cite{Czumaj03perfectlybalanced,Graham:1969}.
Recent studies~\cite{TrafficTypesGrowth:2011,arbor,PADIS2010} show that a small number of content providers are responsible for a large fraction of the traffic.
Therefore it is expected that the algorithm yields results close to the optimal ones.
To further improve the accuracy of the proposed approximation algorithm,
we design an {\it iterative} version of the algorithm, presented in Algorithm~\ref{alg:Greedy-Sort-Flow},
that converges to the optimal solution.
Indeed, a small number of iterations, typically one, suffice to provide
a stable assignment of flows.

As we elaborate in Section~\ref{sec:Evaluation}, we performed a number of simulations using real operational traces,
and different sets of CPs. Our evaluation show that the performance of the iterative greedy algorithm presented in Algorithm~\ref{alg:Greedy-Sort-Flow}
yields results very close to this obtained with LP, but in significantly shorter time.

\section{Active Measurements in ISP1}\label{sec:Active-measurements}

\begin{figure}[t]
 \center\includegraphics[width=1\linewidth]{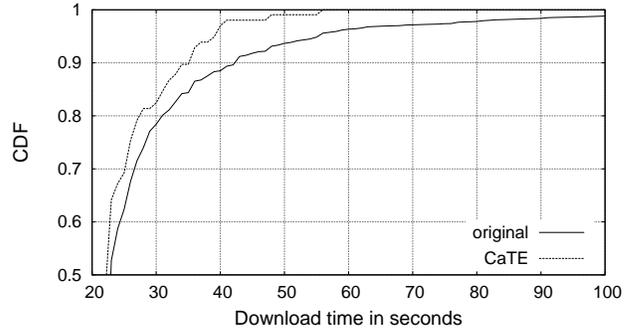}
 \caption{Distribution of download times of a CP.}
 \label{fig:CDF-DLTime}
 \vspace{-1em}
\end{figure}

The \cate evaluation in Section~\ref{sec:CaTE-Main-Results} does not allow us to argue 
about end-user performance, as it is based on simulations. To this end, we complement our 
previous network-wide simulations with active measurements. Over a period of one week, 
we repeatedly downloaded a 60MB object from one of the major CPs. This CP is an OCH 
distributed across 12 locations. The downloads were performed every two hours, from each 
of the 12 locations. Additionally, mapping requests were issued every 200ms to find out the 
dynamics in the server assignment of this CP. Figure~\ref{fig:CDF-DLTime} shows the distribution 
of total download times when the CP assigns end-users to its servers ("original") and compares 
it to the download time that would be observed if \cate had been used. We observe that more 
than $50$\% of the downloads do not show a significant difference. This happens when congestion 
is low, \eg during non-peak hours. For $20$\% of the downloads, we observe a significant difference 
in the download times, mainly during peak hours. 
This confirms our observation 
that \cate is most beneficial during peak hours.

\section{Case study: Netflix in ISP1}\label{sec:Netflix}

\begin{figure}[!t]
\center\includegraphics[width=1\linewidth]{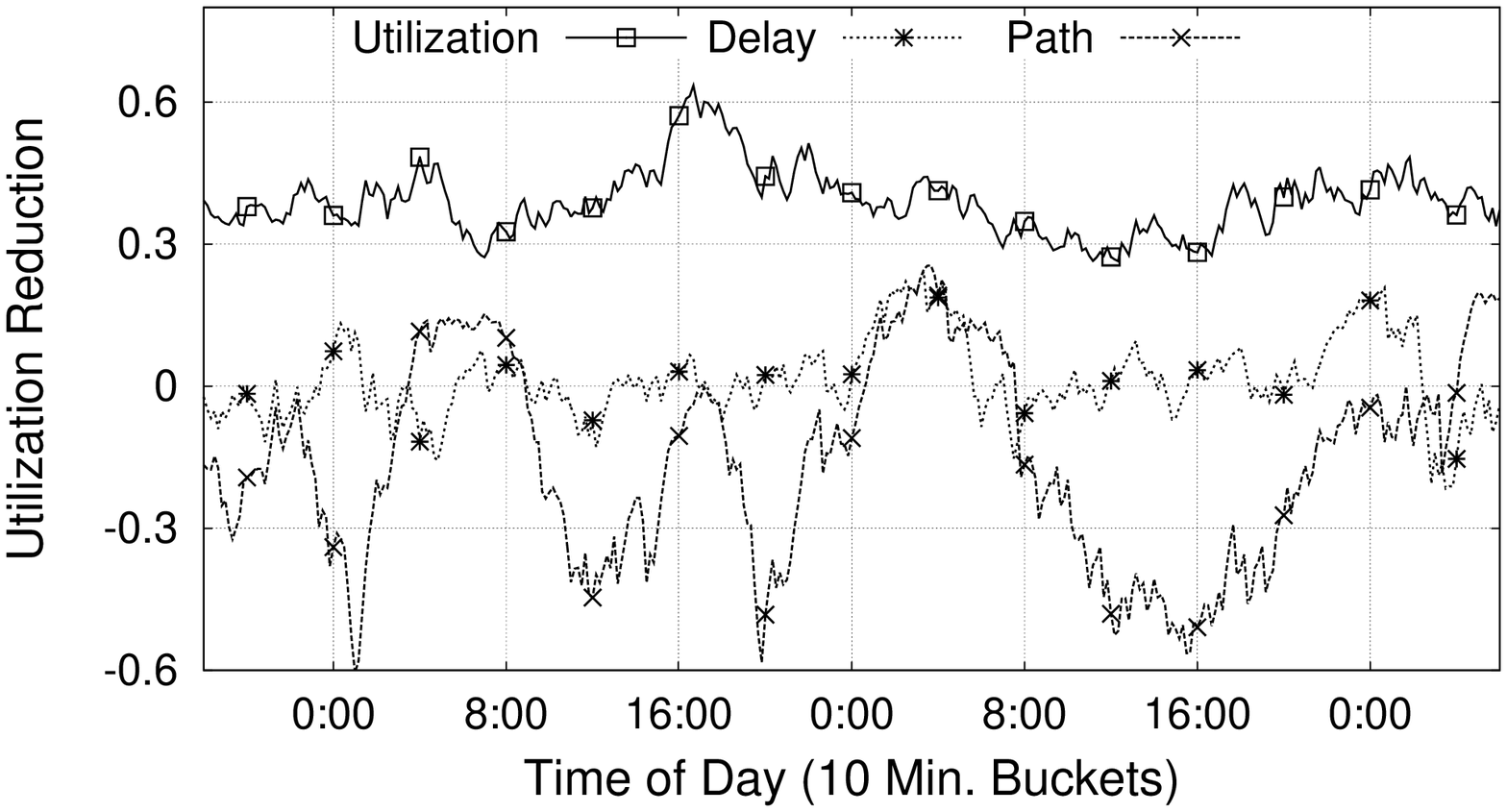}
\center\includegraphics[width=1\linewidth]{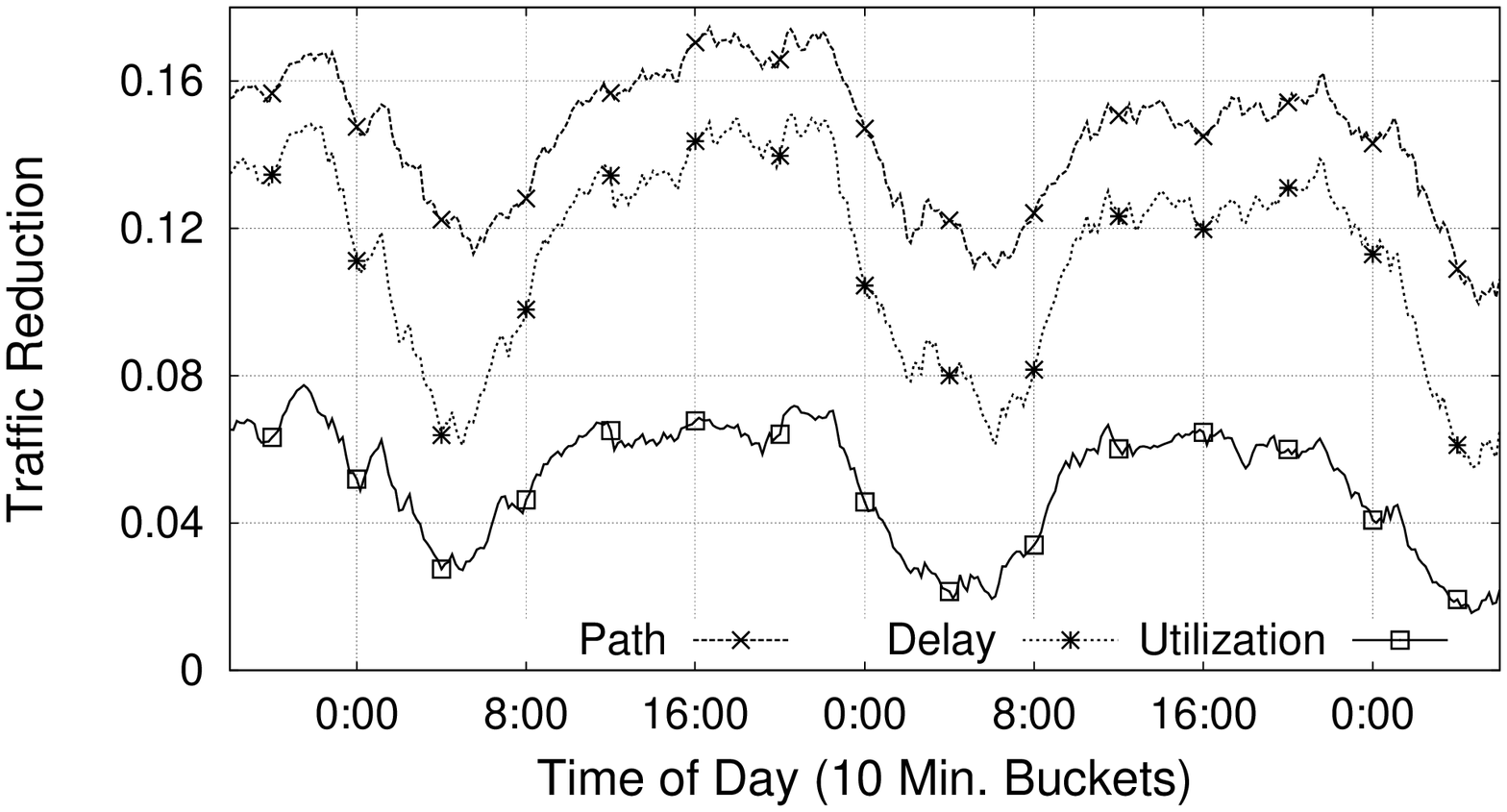}
\center\includegraphics[width=1\linewidth]{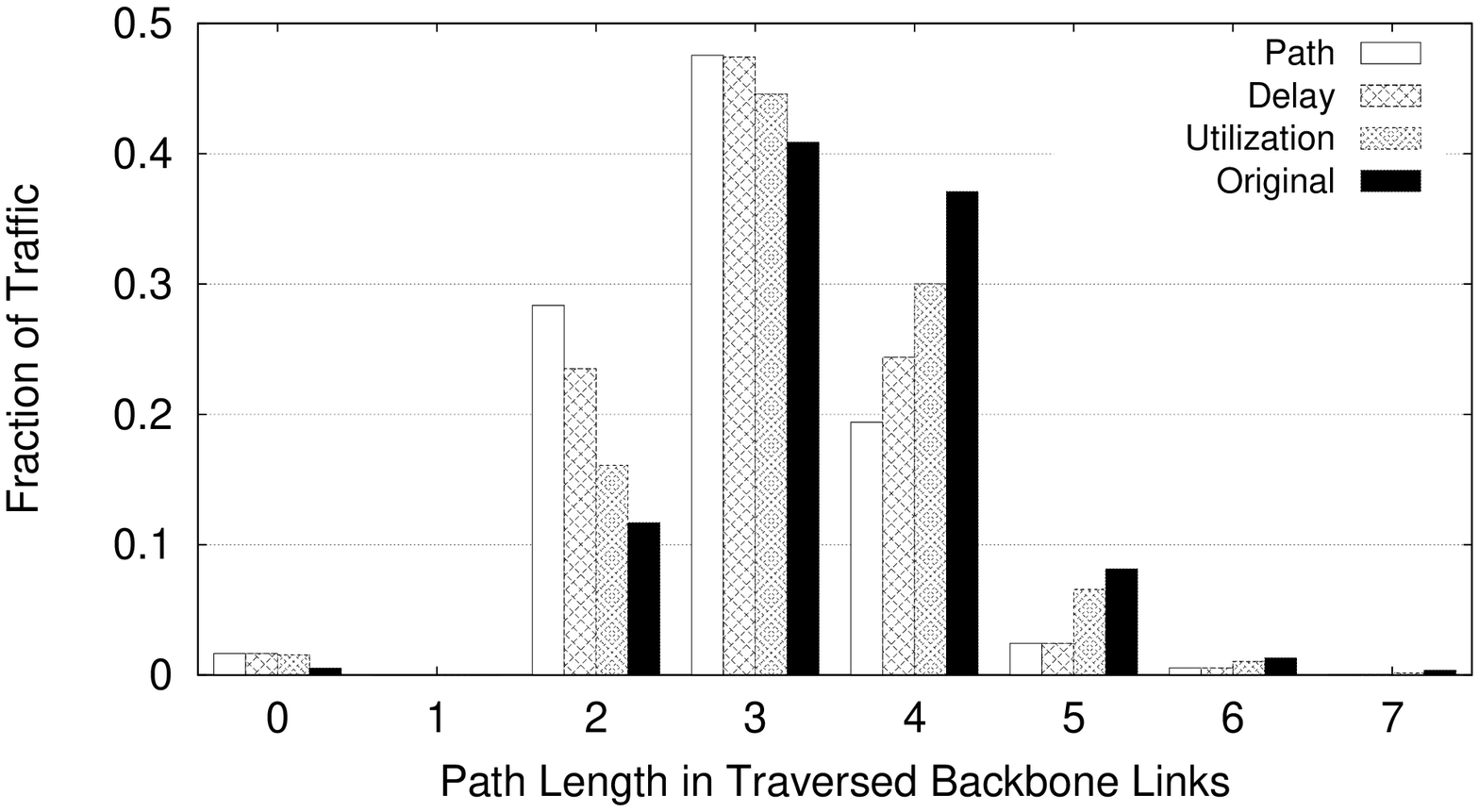}
\caption{Projection of reduction in link utilization (top), reduction in overall network traffic (middle) and 
          fraction of volume by path length (bottom) if Netflix is launched in ISP1.}
\label{fig:Netflix}
\vspace{-1.5em}
\end{figure}

Netflix, a very popular application that delivers high quality videos to end-users, relies on commercial CDNs such 
as Level3 and Limelight to improve the content delivery. Today, Netflix is available in North and Latin America, and 
is announced to arrive in the UK soon. Recent studies show that Netflix is responsible for more than 30\% of the 
peak downstream traffic in large ISPs~\cite{sandvine11}. Consider the scenario where Netflix is launching its 
service in the large European ISP1 we described in Section~\ref{sec:Experimental-Setting}. If the launch happens 
overnight, ISP1 would have to deal with a huge amount of highly variable traffic, which would have significant 
implications on the operation of ISP1. With \cate, the traffic of Netflix can be spread across the ingress points of 
ISP1. This will limit the negative consequences imposed by additional traffic for the CP delivering Netflix as well 
as for ISP1 and thus avoids a deteriorated end-user experience.

To quantify the effect of Netflix being deployed in Europe, we simulate the launch of Netflix in ISP1, assuming 
that the CP currently hosting Netflix increases its traffic 20-fold, while keeping the distribution of the requests.  
Next, we generate a new set of traffic demands for \cate accordingly. We consider the the top 10 CPs by volume 
for \cate, and show the benefits when optimizing for different metrics. 

Our results show that with \cate, the utilization of the most utilized link can be reduced by up to 60\% (see 
top of Figure~\ref{fig:Netflix}), the total HTTP traffic volume can be reduced by 15\% (see middle of 
Figure~\ref{fig:Netflix}) and traffic can be shifted towards shorter paths inside the network of ISP1 
(bottom of Figure~\ref{fig:Netflix}). However, when considering all metrics, we observe that not all metrics 
can be optimized to their full extend at the same time. For example, a reduction of traffic in the order of 15\% would actually 
increase the utilization on the highest loaded link by 60\%. This indicates that the optimization function employed 
by \cate needs to be carefully chosen to target the most important metrics when deploying \cate inside a 
network. Nonetheless, if minimizing the maximum link utilization is chosen as the optimization function for 
\cate, benefits in all metrics can be observed.

Internet applications such as Netflix are in a position to negotiate how they should be deployed in order to 
improve end-user experience and not disturb the operation of ISPs. \cate can be used to identify the best 
peering points between the CPs that deliver Netflix traffic and the ISPs that receive its traffic. In addition, 
ISPs might offer better peering prices if the CPs hosting Netflix are willing to provide a higher diversity in 
the locations from which the traffic can be obtained. This would lead to a win-win situation where Netflix 
can offer better service to its users, the CPs achieve reduced pricing on their peering agreements, and ISPs 
can compensate the reduced peering revenue through more efficient operations.
\end{appendix}
\end{document}